\documentclass[journal,article,submit,moreauthors,pdftex]{Definitions/mdpi} 
\usepackage{placeins}
\firstpage{1} 
\pubvolume{1}
\issuenum{1}
\articlenumber{0}
\pubyear{2021}
\copyrightyear{2020}
\datereceived{} 
\dateaccepted{} 
\datepublished{} 
\hreflink{https://doi.org/} 

\newcommand\aap{A\&A}                
\newcommand\aaps{A\&AS}              
\newcommand\aj{AJ}                   
\newcommand\apj{ApJ}                 
\newcommand\apjl{ApJ}                
\newcommand\apjs{ApJS}               
\newcommand\araa{ARA\&A}             
\newcommand\mnras{MNRAS}             
\newcommand\nat{Nature}              
\newcommand\pra{Phys. Rev.~A}        
\newcommand\prl{Phys. Rev.~Lett.}    
\newcommand\pasp{PASP}               

\def\blankline{\par\vskip 12pt\noindent}
\def\blankhalf{\par\vskip 6pt\noindent}

\newcommand\ion[2]{#1\,\,{\sc{\romannumeral #2}}}
\newcommand{\kms}{\hbox{km\,s$^{-1}$}}

\newcommand{\Vinf}{\hbox{$V_\infty$}}

\newcommand{\Jbar}{\hbox{$\bar J$}}
\newcommand{\Rstar}{\hbox{$R_*$}}
\newcommand{\Lstar}{\hbox{$L_*$}}
\newcommand{\Rsun}{\hbox{$R_\odot$}}
\newcommand{\Lsun}{\hbox{$L_\odot$}}
\newcommand{\Msun}{\hbox{$M_\odot$}}
\newcommand{\Mstar}{\hbox{$M_*$}}
\newcommand{\Msunyr}{\hbox{$M_\odot \,\hbox{yr}^{-1}$}}
\newcommand{\Mdot}{\hbox{$\dot M$}}
\newcommand{\Teff}{\hbox{$T_{\scriptstyle \rm eff}$}}

\newcommand{\chir}{\hbox{$\chi_{\scriptstyle R}$}}

\newcommand{\tsup}[1]{\textsuperscript{#1}}
\newcommand{\tsub}[1]{\textsubscript{#1}}

\newcommand{\emin}{\hbox{$e^{-}$}}

\newcommand{\cpd}{CPD--56$^{\rm o}$~8032}
\newcommand{\cmfgen}{{\sc cmfgen}}

\newcommand{\civres}{\ion{C}{4}\ $\lambda\lambda 1548, 1551$}
\newcommand{\nvres}{\ion{N}{5}\ $\lambda\lambda 1238, 1243$}
\newcommand{\ovires}{\ion{O}{6}\ $\lambda\lambda 1032, 1038$}

\newcommand{\ssb}{\hbox{$\sigma_{\scriptstyle \rm SB}$}}
\newcommand{\odd}[2]{\hbox{$^#1${\rm #2}$^{\scriptstyle \rm o}$}}
\newcommand{\even}[2]{\hbox{$^#1${\rm #2}}}

\Title{Photoionization and electron-ion recombination in astrophysical plasmas}

\TitleCitation{Title}

\Author{D. John Hillier $^{1}$\orcidA{0000-0001-5094-8017}$^*$}

\AuthorNames{D. John Hillier}

\AuthorCitation{Hillier, D. John}

\address{
$^{1}$ \quad Department of Physics and Astronomy \& Pittsburgh Particle Physics, Astrophysics and Cosmology Center (PITT PACC), University of Pittsburgh, 3941 O'Hara Street,  Pittsburgh, PA 15260, USA; hillier@pitt.edu.c}

\corres{Correspondence: hillier@pitt.edu; Tel.:  +1 412-624-9000}

\abstract{Photoionization and its inverse, electron-ion recombination, are key processes that
influence many astrophysical plasmas (and gasses), and the diagnostics that we use to analyse the plasmas. In this review we provide a brief overview of the importance of photoionization and recombination in astrophysics. We highlight how the data needed for spectral analyses, and the  required accuracy, varies considerably in different astrophysical environments. We then discuss photoionization processes, highlighting resonances in their cross-sections. Next we discuss radiative recombination, and low and high  temperature dielectronic recombination. The possible suppression of low temperature dielectronic recombination (LTDR) and high temperature dielectronic recombination (HTDR) due to the radiation field and high densities is discussed. Finally we discuss a few astrophysical examples to highlight photoionization and recombination processes.}


\begin{document}

\end{paracol}
\section{Introduction}

 One of the most important ways we learn about the Universe is through spectroscopy. From spectroscopy we can typically deduce important stellar parameters such as a star's effective temperature,\footnote{The effective temperature of the star is defined by the relation $L=4\pi R_*^2 \ssb \Teff^4$ where $L$ is the stellar luminosity (energy emitted per second), \ssb\ the Stefan-Boltzmann constant) and $R_*$ is the radius of the star} surface gravity ($=G\Mstar/\Rstar^2$) and abundances. These in turn provide insights into stellar evolution, galactic evolution, and the evolution of the Universe. To perform analyses of stellar data
requires atomic data, although the amount and type of atomic data needed varies greatly with the application. In the most extreme cases, in which local thermodynamic equilibrium does not hold (discussed below), we require, for example, photoionization cross-sections, oscillator strengths for  bound-bound transitions, line-broadening data, collisional cross-sections, autoionization rates, chemical reaction rates, and charge exchange cross-sections. For some simple species, such as hydrogen, we have excellent atomic data while for other important species, such as Fe group elements, crucial atomic data is lacking. Unfortunately, for many ionization stages even basic information, such as accurate energy levels, is also missing.\footnote{
The meaning of ``accurate" is highly context dependent. Atomic data calculations can, in some cases, give energy levels accurate to 1\%, and for some purposes this is sufficient. However, for spectral modeling such energy levels cannot be used to compute transition wavelengths --  a 1\% shift (which will be potentially larger if the levels are close in energy)  will move a line way far from its correct location, influencing spectral synthesis calculations. Moreover, in non-LTE a wrong wavelength will influence how a line interacts with neighboring transitions. In O supergiants two weak  \ion{Fe}{4}\ lines, that overlap with
\ion{He}{1} $\lambda$304, influence the strength of \ion{He}{1} singlet transitions in the optical, and accurate
wavelengths (and oscillator strengths) of these \ion{Fe}{4}\ lines are crucial for understanding the  \ion{He}{1} singlet transitions \citep{NHP06_HeI}.} However, invaluable work by the NIST and Imperial college atomic spectroscopy groups is helping to rectify this situation for some important astrophysical ions \citep[e.g.,][]{2013ApJS..204....1N,2022ApJS..261...35C}.

In this review we discuss the importance of photoionization cross-sections for astronomical applications, with an emphasis on massive stars. Such a discussion will necessarily consider recombination processes -- the inverse of photoionization processes. Before doing so it is necessary to define some important physical and astronomical terms.

A star (or any other astrophysical object emitting radiation) is not in thermal equilibrium. However in some cases, and at some locations, it is well justified to assume that the plasma in the star is in ``local" thermal equilibrium. Below the atmosphere (the thin layer that emits the radiation we observe) the material in most stars can be considered to be in ``local" thermal equilibrium (LTE). In such cases, the state of the gas (e.g., the thermodynamical properties, the ionization state, and the populations of atomic levels) are set by the density and electron temperature via thermodynamic arguments. The ionization state of the gas and the level populations are determined, for example, by the Saha and Boltzmann equations  \citep[e.g.,][]{Mih78_book}. Moreover, there is only one temperature --- the electron temperature, the ion temperature, the excitation temperature, and the radiation temperature are all identical.  The temperature of the gas varies with location (it must, since radiation is propagating outwards), but the scale on which it varies does not affect the thermodynamic state of the gas. Unfortunately, much of the radiation we observe comes from gas that is NOT in LTE -- typically referred to as nLTE or non LTE.

At the stellar surface the assumption of LTE becomes less valid. This is not surprising -- at the stellar surface radiation is escaping from the star and there is no incident radiation (at least for single stars), and hence the radiation density must drop from its blackbody value by a factor of (at least) $\sim 2$ (since there is
no incident radiation -- see \cite[see][]{Mih78_book}, p.~120). Further, because radiation can now travel significant distances, regions of different temperatures are directly coupled, potentially making the radiation field at a given location strongly non-Planckian.

Fortunately, in some cases the densities are high enough that collisional processes can still strongly couple the level populations  and the ionization state of the gas to the local electron temperature, allowing us to use the Saha and Boltzmann equations to compute level populations. In such cases the electron temperature (which will be the same as the ion temperature) determines the state of the gas, and it is this temperature that we normally state. In general, however, the electron and radiation temperatures will be different.\footnote{Strictly speaking, a radiation temperature is only well defined if the
radiation field is Planckian. However, astronomers  often use color temperatures, defined by fitting a scaled
blackbody ratio to the flux at two wavelengths, to characterize the nature of the radiation field in some pass band. In nLTE, astronomers may also use the excitation temperature to characterize the excitation or ionization state of a gas. In general these will not be the same as the local electron temperature, and will vary
with level and ionization stage (though possibly in a systematic way).}

The Boltzmann equation, which relates the populations of two levels within the same ionization state, is
  \begin{equation}
      n_u^* =  {g_u  n_l^* \over g_l} exp{(-E_{lu}/kT)}
   \end{equation}
where $n_l^*$ and $n_u^*$ are the  LTE population densities of the lower and upper states respectively,
$g_l$ and $g_u$ are the level degeneracies,  and $E_{lu}$ is the difference energy between the two levels
\citep[e.g.,][]{Mih78_book}.

The Saha equation, which relates the ground state populations of two consecutive  ionization equations, is
\begin{equation}
   n^*_{1,i} =n^*_{1,i+1} N_e { g_{1,i} C_I \over g_{1,i+1} T^{3/2} } \exp{(-\psi_i/kT)}
\end{equation}
where T is the electron temperature in Kelvin, $C_I= 2.07 \times 10^{-16}$ (cgs units), $N_e$ is the electron density, $\psi_i$ is the ionization energy, and the subscript
$i$ ($i+1$) is used to denote the ionization stage. 

\begin{figure}[htbp]
\centering
\includegraphics[width=8.0 cm]{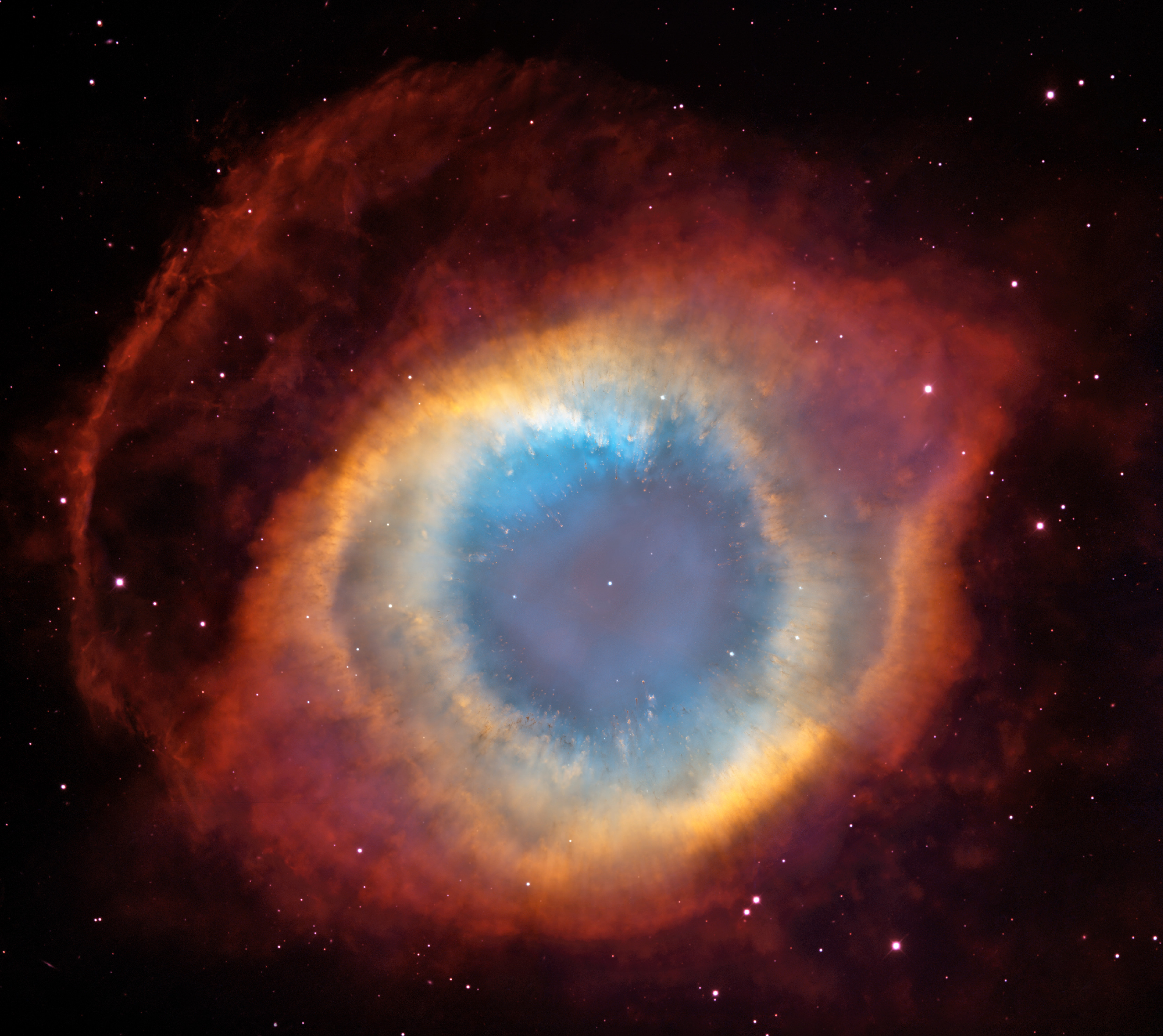}
\caption{A composite image of the Helix nebula obtained using the HST and the Cerro Tololo Inter-American Observatory in Chile. H$\alpha$+[\ion{N}{2}] ($\lambda\lambda 6548.1, 6583.4\,$\AA), emission is shown in red, an average of H$\alpha$+[\ion{N}{2}] and [\ion{O}{3}] ($\lambda\lambda 4958.9, 5006.8\,$\AA) is shown in green, and forbidden  [\ion{O}{3}] emission is shown in blue \citep{2004AJ....128.2339O}.
The nebula lies at a distance of about 200 pc, and the size of the
semi-major axis is approximately 5.5 arc minutes or  66,000 AU \citep{2004AJ....128.2339O}. 
 The central star has an effective temperature of 104,000\,K, $L \approx 80$\,\Lsun, and $R\approx 0.028\,\Rsun$ \citep{2009AJ....138.1969B}. Using the semi-major axis as a representative
size scale, the dilution factor (\,$0.25(r/\Rstar)^2$\,) is of order $10^{-18}$.  Detailed insights into
the structure and morphology of the Helix nebula  can be found in many works \citep[e.g.,][]{2004AJ....128.2339O,2008MNRAS.384..497M}. \\
Image credit: NASA, ESA, C.R. O'Dell (Vanderbilt University), and M. Meixner, P. McCullough, and G. Bacon (Space Telescope Science Institute).
}
\label{fig_helix}
\end{figure}

In many stars, and especially those with lower density gas (nebulae, stellar winds, supernovae) the departures from LTE are significant and MUST be allowed for. Gaseous nebula, which typically have densities less than $10^6$ atoms\,cm$^{-3}$ provide an excellent example in which the departures from LTE are extreme. The radiation field that ionizes the
nebula typically emanates from a hot star ($\Teff \gtrsim 25,000\,$K) and is strongly diluted since
the nebula is typically very distant from the star (Figure~\ref{fig_helix}). The equilibrium temperature of the gas is typically around 10,000\,K (which is primarily determined by the chemical composition of the nebula), and is insensitive to the effective temperature of the star.  The nebula,  longward (i.e., at larger wavelengths)  of the \ion{H}{1}\ Lyman jump at 912\,\AA, is transparent to most radiation.  Because of the non-Planckian radiation field and the low densities, collisions with electrons cannot drive the gas into LTE.

When LTE no longer holds (i.e., nLTE)  we are forced to solve for the ionization state of the gas and the level populations from first principles. That is, we need to consider all the
processes, and inverse processes, that populate a given level. These processes include
photoionization and recombination, bound-bound emission and absorption, collisional excitation and de-excitation, collisional ionization and collisional recombination, dielectronic recombination and autoionization, charge exchange reactions and, in  ``cooler gas" ($T \lessapprox 6000$\,K) chemical reactions, and dust chemistry.\footnote{The Sun's atmosphere is cool and dense enough for molecules to form, and 50 molecular
species have been identified \citep{2003ASPC..288..303J}.  In the solar spectrum, spectral features due, for example, to CO, SiO, H$_2$, OH, CH, C$_2$, and CN, have been identified \citep{1994LNP...428..196G,2003ASPC..288..303J}. Dust formation in red giants and supergiants is common but not well understood \citep{2017ASPC..508...57C}, and in some cases may be associated with non-equilibrium chemistry \citep{2016A&A...585A...6G}.} Determining the state of the gas is a complicated problem. Many of the processes above depend on the radiation field which in turn depends on the level populations. Further, the radiation field couples the gas to regions of different temperatures. This is a highly non-linear problem, and can only be solved by iterative techniques. In many cases we can consider the population numbers to be static (and the equations are referred to as the equations of statistical equilibrium), but in other cases (e.g., supernovae) we may need to allow for time dependence (and solve the kinetic equations). Another major issue is correcting for plasma effects that limit the number of levels in atoms and ions (i.e., crudely,
an atom/ion cannot be larger than the inter-atom spacing). One approach
 is  probabilistic and was developed in a series of papers by Hummer, Mihalas, and Dappen \citep{1988ApJ...331..794H,1988ApJ...331..815M,1988ApJ...332..261D},
 and is used, for example, in the nLTE radiative transfer codes {\sc tlusty}\citep{HL95_TLUSTY} and {\sc cmfgen} \citep{HM98_blank}. In this approach, levels are assigned an occupation probability that varies smoothly with the level energy, density and temperature, and that leads to a finite partition function.

Thus a vast amount of atomic data is needed. Sadly, despite heroic efforts by Bob Kurucz \citep{1995KurCD..23.....K, Kur09_ATD}, members of the Opacity Project \citep{Sea87_OP}
and Iron Project \citep{1993A&A...279..298H}, Sultana Nahar  \citep{2020Atoms...8...68N} and many others, much of the needed data is still missing. Extensive photoionization data is available, for example, through
\href{http://cdsweb.u-strasbg.fr/topbase/topbase.html}{TOPbase}
 \citep{Topbase93}, \href{http://cdsweb.u-strasbg.fr/tipbase/home.html}{TIPbase} \citep{2000AIPC..543..313M} and \href{https://norad.astronomy.osu.edu}{NORAD}  \citep{2020Atoms...8...68N}. As discussed later in this article, details can matter, and it is not always obvious, a priori, which data is essential for accurate analyses. As this paper is concerned with  photoionization/recombination, this paper does not generally elaborate on other important processes.
Information on these additional processes can be found in \cite[e.g.,][]{1986rpa..book.....R,2006agna.book.....O,2015aas..book.....P,2016mmcr.book.....R}.

 Below we discuss photoionization and recombination as relevant to astrophysical
applications. Much of the following discussion will be based on my own experiences in developing \cmfgen, a nLTE radiative transfer code originally designed to model hot massive stars $(M> 20\,\Msun, \Teff \ge 20,000\,$K) and their stellar winds \citep{H90_lin,HM98_blank}.  The winds in these stars are driven by radiation pressure acting on bound-bound transitions belonging, for example, to C, N, O, Ar, and Fe
\citep[e.g.,][]{CAK75,PPK86_winds,2019A&A...632A.126S}. Since its  initial development the code has undergone considerable revisions and improvements. It has been successfully used to model O stars \cite[e.g.,][]{MSH02_Teff,CHE02_FUSE,BLH03_ngc340,HLH03_AV83,2012A&A...544A..67B}, Wolf-Rayet (WR) stars \citep{2021MNRAS.503.2726H,2022ApJ...931..157A}, luminous blue variables \citep[e.g.,][]{2001ASPC..233..133N,2009ApJ...698.1698G,GHD11_AG_CAR}, B stars \citep[e.g.,][]{2016MNRAS.456.2907P}, the central stars of planetary nebulae \citep[e.g.,][]{HB04_cspn}, and A stars. Over the last decade \cmfgen\ was adapted to  treat  time-dependent radiation transfer, and to solve the time-dependent kinetic equations \citep{HD12_time}, and it has been used to model spectra resulting from a variety of SN explosions \cite[e.g.,][]{2020MNRAS.494.2221W,2020A&A...642A..33D,2021A&A...652A..64D,2021A&A...656A..61D}.

The review is organized as follows: In \S\ref{Sect_interior} we briefly discuss the importance of
photoionization processes for stellar interiors and introduce the Rosseland mean opacity. We then consider photoionization processes in \S\ref{Sect_phot} with an emphasis on inner shell ionizations in \S\ref{Sect_in_shell}. Recombination processes are discussed in
\S\ref{Sect_recom} while suppression of dielectronic recombination by collisions
and the radiation field is discussed in \S\ref{Sect_sup_col} and \S\ref{Sect_sup_rad}
respectively. Specific examples of where photoionization/recombination processes
are important are then discussed -- direct recombination (\S\ref{Sect_ex_recom}), the Sun  (\S\ref{Sect_sun}),
O stars, WR stars, luminous blue variables (LBVs),  (\S\ref{Sect_OWR}),  Of and WN stars  (\S\ref{Sect_OfWN}), carbon lines in WC stars  (\S\ref{Sect_C_WC}), \ion{C}{2} in a [WC] star  (\S\ref{Sect_C2_WC}), and supernovae  (\S\ref{Sect_SN}).\

\section{Stellar Interiors}
\label{Sect_interior}

In stellar interiors energy is transported by radiation and by convection, and in degenerate
stars, by conduction. As LTE holds, photoionization processes have no direct influence on the level populations (since the populations are determined by the Saha and Boltzmann equations), but they do help to determine the temperature of the gas, and they do help to set the continuous radiation field that we observe. Due to the small mean-free-path of photons, radiation transport is diffusive, and in this regime a single quantity, the Rosseland mean opacity, is required to describe the transport of radiative energy. At depth in the star the radiation diffuses, and the specific intensity ($I_\nu$) at frequency $\nu$ is given
by
\begin{equation}
 I_\nu = B_\nu - {\mu \over \chi_\nu} {dB_\nu \over dr}
 \end{equation}
\citep[e.g.,][]{Mih78_book} where $B_\nu$ is the Planck (i.e., blackbody) function
given by
\begin{equation}
B_\nu = {2 h\nu^3 \over c^2}{1 \over \exp(h\nu/kT)-1}\,\,\,,
\end{equation}
$\chi_\nu$ is the opacity, and $\mu$ is the angle between the radius and the direction of radiation propagation. The radiative flux is simply given by
\begin{equation}
F_\nu= {-4\pi\over  3\chi_\nu} {dB_\nu \over dr}\,\,
\end{equation}
The negatives sign in the expression for the flux arises because $T$, and hence $B_\nu$, decrease with increasing $r$. Integrating over all frequencies yields a total radiative flux, $F$, given by
\begin{equation}
F=  -16 \sigma {T^3 \over  3 \chir}  {dT\over dr}
\end{equation}
where $\sigma$ is the Stefan-Boltzmann constant and $\chir$ is the Rosseland mean opacity as defined by
\begin{equation}
{1\over \chir} = { \int_0^\infty  {1 \over \chi_\nu} dB_\nu/dT \, d\nu \over \int_0^\infty  dB_\nu/dT \,\ d\nu}\,\,.
\end{equation}

\noindent
In stellar interiors the Rosseland mean opacity is a function of density, temperature, and composition, and is primarily determined by the most abundant species, with H, He, C, N, O, Ne, and Fe being the most crucial. As it is a harmonic mean it can be strongly influenced by regions of low opacity. Thus it is crucial to take into account all opacity sources, particularly contributions in regions of otherwise low opacity. For a given species, the opacity is determined by photo-ionization processes, bound-bound transitions, and free-free processes. The required cross-sections are non trivial to compute, especially for Fe group elements (with a partial filled 3d shell). 
Fortunately, because it is a broad integral, the Rosseland mean opacity is insensitive to small random errors (e.g., bound-bound transitions slightly offset from their correct wavelength) in the atomic data.


The computation of the Rosseland mean opacity is further complicated by the need to account for plasma effects -- atoms/ions in a star are not isolated, but  experience a time-varying electric field due to their neighbors. This broadens bound-bound transitions which enhances the Rosseland mean opacity since the influence of the line is spread over a broader band into regions which may have lower opacity. Further, the size of the atoms/ions (and hence the number of levels) will be truncated since atoms/ions can only occupy a finite volume. The latter can be thought of as a lowering of the ionization potential, but more rigorous approaches, for example, use probabilistic arguments \citep{1988ApJ...331..794H,1994A&A...282..151H}. An extensive discussion of some of the issues related to plasma effects on the equation of state, and additional references, are given by \cite{2021LRSP...18....2C}. Extensive efforts have been made to provide LTE opacity libraries for stellar astrophysics that take into account different physical effects with varying degrees of fidelity. These
include the Opacity Project \citep{1994MNRAS.266..805S}, opacities computed using the {\sc opal} code \citep[e.g.,][]{1996ApJ...464..943I,2016ApJ...817..116C}, the {\sc opas} code \citep{2012ApJ...745...10B}, 
and a suite of codes developed at The Los Alamos National Laboratory \citep{1995ASPC...78...51M}.

 
 
 

\section{Photoionization}
\label{Sect_phot}

The photoionization rate from a level $l$ (in an arbitrary ion of arbitrary charge) can be written as
\begin{equation}
               \left(dn_l\over dt\right)_{\scriptsize \rm PR}= - n_l \int_{\nu_o}^\infty \left({4\pi \over h\nu}\right) \sigma_\nu J_\nu d\nu
\end{equation}

\noindent
where $n_l$ is the population density of level $l$, $\sigma_\nu$ is the frequency dependent
photoionization cross-section (units are cm$^{2}$), $\nu_o$ the threshold frequency for ionization, $h$ is
Planck's constant,  and $J_\nu$ is the mean intensity
(erg\,cm$^{-2}$\,s$^{-1}$\,Hz$^{-1}$). $J_\nu$
is defined by
\begin{equation}
J_\nu = {1\over 4\pi} \oint I_\nu \,d\Omega
\end{equation}
where $d\Omega$ is an increment in solid angle. If the radiation field is Planckian and isotropic,
\begin{equation}
J_\nu=I_\nu=B_\nu\,\,.
\end{equation}

The photoionization cross-sections are generally obtained from numerical calculations -- it
is not feasible to measure all the cross-sections in the laboratory. Rather, laboratory measurements
are  used to test the accuracy of theoretical calculations. The accuracy of the cross-section varies
greatly being dependent on both assumptions used in the modeling, and on the complexity of the model atom.
For example, it is much easier to compute atomic data for atoms/ions with a partially filled  p shell,
than it is for the lanthanides and actinides which have a partially filled 4f or 5f shell respectively \cite[e.g.,][]{2012PhRvA..85e2716B}. The lanthanides are believed to be created in neutron-neutron star mergers, and are an important opacity source in the outburst spectra that result from such mergers \citep{2010MNRAS.406.2650M,2013ApJ...774...25K,2013ApJ...775...18B,2020MNRAS.493.4143F,2020MNRAS.496.1369T,2023MNRAS.519.2862F}.

\noindent
Typically in model atmosphere codes the rates (integrals) are evaluated using numerical quadrature. Thus
\begin{equation}
\left(dn_l\over dt\right)_{\scriptsize \rm PR}= - n_l \sum_i \left({4\pi \over h\nu_i}\right) w_i \sigma_i J_i 
 \end{equation}
where $w_i$ is the quadrature weight at frequency $\nu_i$. 

In a simple species such as hydrogen, the photoionization process is simply 
\blankhalf
\indent
         H + $h\nu  \rightarrow$ H$^+$ + e$^-$. 
\blankhalf
\noindent
In more electron-rich species the process is more complicated
since there are multiple photoionization routes.
For example,  there are two direct photoionization routes from \ion{C}{3}(2s 2p): 
\blankhalf
\indent
\ion{C}{3}(2s 2p \tsup{1}P\tsup{o}) + $h\nu  \rightarrow$ \ion{C}{4}(2s $^2$S) + \emin\ \,\,\,\, \& 
\blankhalf
\indent
\ion{C}{3}(2s 2p $^1$P\tsup{o}) + $h\nu \rightarrow$ \ion{C}{4}(2p $^2$P\tsup{o}) + \emin 

\blankhalf
\noindent
The first process occurs provided the photon energy exceeds the ionization energy of 
the \ion{C}{3}(2s 2p \tsup{1}P\tsup{o}) state\footnote{Throughout the article we neglect full shells when providing the electron configuration. We use the principal quantum number ($n$), orbital angular momentum number ($l =0,\dots,n-1 = $\,\,s, p, d, f, g\,$\dots$), and spin  ($\pm 1/2$) to describe the state of an electron. Thus 2p indicates an electron with $n=2$ and $l=1$. LS-coupling (in which the orbital angular momenta are coupled and the electron spins are coupled) is used to provide the term designation. A term designation has the format \tsup{2S+1}L\tsup{x} where S is the sum of the (valence) electron spins, L is the total orbital angular momentum, and ''o" is used to indicate that the arithmetic sum of the electron orbital angular momenta is odd (o) or even (in which case e is omitted by convention). An excellent primer on atomic spectroscopy is provided by \citep{nist_primer}.}. The second process occurs when the photon energy exceeds
the sum of the ionization energy and the difference in energy between the 2s and 2p states in \ion{C}{4}. Of course, photons of sufficient energy may also ionize C$^{2+}$ by ejecting an inner (1s) electron  -- a process of great importance when X-rays are present.

\begin{figure}[htbp]
\centering
\includegraphics[width=12.0 cm]{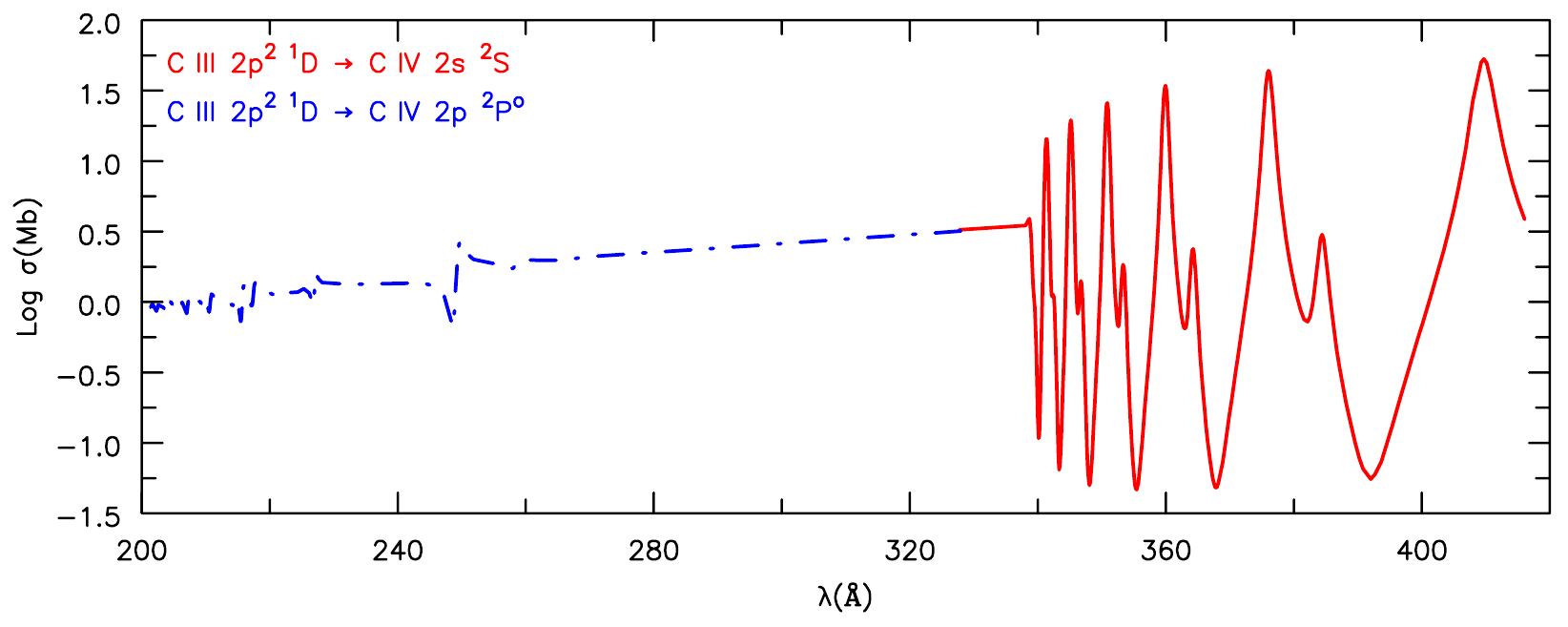}
\caption{Illustration of the photoionization cross-section (in megabarn, with 1\,Mb\,$ = 10^{-18}$ cm$^{2}$) of C\,{\sc iii} 2p$^2$\,\even{1}{D}. Ionization to the \ion{C}{4} ground state occurs via autoionizing levels such as
\ion{C}{3} 2p 4d \odd{1}{F}. Shortward of $\approx$325\,\AA, photons have sufficient energy to
ionize directly to \ion{C}{4} 2p\,\odd{2}{P}. The data were convolved with a Gaussian profile with a full-width at
half maximum of $\sim 600$\,\kms (i.e., $\sigma=250.0$\,\kms). The photoionization cross-sections for \ion{C}{3} were computed by P. J. Storey (private communication). In other photoionization cross-sections the resonances are often much narrower than those shown here.
}
\label{fig_ciii_2p2}
\end{figure}

There may also be multiple indirect photoionization routes such as: 
\blankhalf
\indent
\ion{C}{3}(2p$^2$ $^1$D) + $h\nu  \rightarrow$ \ion{C}{3} (2p 4d $^1$F\tsup{o})  $\rightarrow$ \ion{C}{4}(2s $^2$S) + \emin 
\blankhalf
The above produces a relatively  ``narrow" resonance in the photoionization cross-section -- it is narrow since the photon has to have the right energy to excite one of the 2p electrons into the 4d state (Figure~\ref{fig_ciii_2p2}). The energy of this state lies above the \ion{C}{4} ground state. The last step in this process is referred to as autoionization. 

In LTE the final state arising from the photoionization process is irrelevant -- only the total opacity matters.  In general, in nLTE, the final state matters, since each process contributes to the population of a different state whose population needs to be determined from first principles. In practice this is generally not a crucial concern for most spectral modelling since the rates for  processes connecting states within an ion are generally much larger than the photoionization and recombination rates.  However, there are cases where the final-state-dependent cross-sections are important.

As a first example, we again consider the \ion{C}{3} / \ion{C}{4} system in WR stars\footnote{WR stars are a class of massive stars that evolved from O stars (stars with initial masses $\gtrsim$15\,\Msun). They are experiencing mass loss via a stellar wind (induced by radiation pressure acting through bound-bound transitions) with a mass-loss rate typically in excess of $10^{-5}$\,\Msunyr, and a terminal wind speed of
$\sim$1000 to 3500\,\kms\ \cite{2000eaa..bookE1895H,2007ARA&A..45..177C}.  In many WR stars the wind is sufficiently dense that the entire spectrum we observe originates in the wind -- the hydrostatic core of the star is not seen. There are two main WR classes: WN stars exhibit  N and He (and sometimes H) emission lines,  and exhibit enhanced N and He at the stellar surface due to the CNO cycle (the main H-fusion chain in massive stars). WC stars exhibit emission lines of He, C, and O, with a C abundance comparable to that of He \cite[e.g.,][]{SHT12_GAL_WC,2022ApJ...924...44A}. They have lost all of their hydrogen envelope, with the enhanced C abundance arising from the triple alpha process (3\,$^4$He$\rightarrow ^{12}$C).}
In the photosphere the strong \civres\ doublet (due to 1$s^2$\,2s--1$s^2$\,2p) is optically thick, and hence the 1$s^2$\,2p state is strongly coupled to the 1$s^2$\,2s ground state through collisional de-excitation and excitation (i.e., the 2p state is in LTE (computed using the local electron temperature) with respect to the 2s state ). Consequently we can treat all ionizations/recombinations as occurring to/from the \ion{C}{4} ground state. However, as the density declines photon escape in the resonance line will lead to a decoupling of 1$s^2$\,2p from the 1$s^2$\,2s state, in which case we should treat recombinations from the 1$s^2$\,2p state separately from those occurring from the 1$s^2$\,2s state. Fortunately, because the 1$s^2$\,2p state lies  $\approx 8$ eV above the ground state,  recombinations from the 1$s^2$\,2p state are generally not very important for the temperatures and densities appropriate to WR stars in the regime important for spectrum formation. This may not be the case in other regimes,
and for other ions with states closer in energy to the ground state.

One crucial area where state-dependent photoionization cross-sections are important
is in X-ray fluorescence where the ejection of an inner shell electron leads to the ion being in a highly excited state, and the emission of characteristic X-rays or the subsequent ejection of one or more additional electrons (Auger ionization) \citep[e.g.,][]{Wei74_Xray_cross,KM3_Auger}. The subsequent decay of these more highly charged ion gives rise to lines which can be detected, and their strength is
dependent on the details of the autoionization processes that occurred after the
inner shell electron was ejected \citep[e.g.,][]{1969PhRv..185....1M,1970PhRvA...2..273M,1972RvMP...44..716B}.

\subsection{Inner shell ionization}
\label{Sect_in_shell}

Typically ionization from the inner shell of an ion (e.g., from the 1$s^2$ shell in \ion{O}{1}--\ion{O}{5}) is
not very important for modelling stellar spectra, since very little flux will be emitted at the relevant energies when 
the ionization stages are abundant. An exception occurs when there is a significant source of X-rays, as can occur when a star has a corona, or when there is a compact object with an accretion disk. For massive stars, X-rays can arise in shocks generated by a wind-wind collision in a binary system, or in shocks generated
from instabilities in the driving of the wind by radiation pressure \citep[e.g.,][]{OCR88,1995A&A...299..523F,2018A&A...611A..17S}. For O stars, the observed X-ray fluxes generated by these two processes are typically in the range of $10^{-5}$ to $10^{-8}$\,\Lstar\  \citep[e.g.,][]{Chl89_xrays,BSC96_OB_rosat}.

\begin{figure}[htbp]
\centering
\includegraphics[width=7.5 cm]{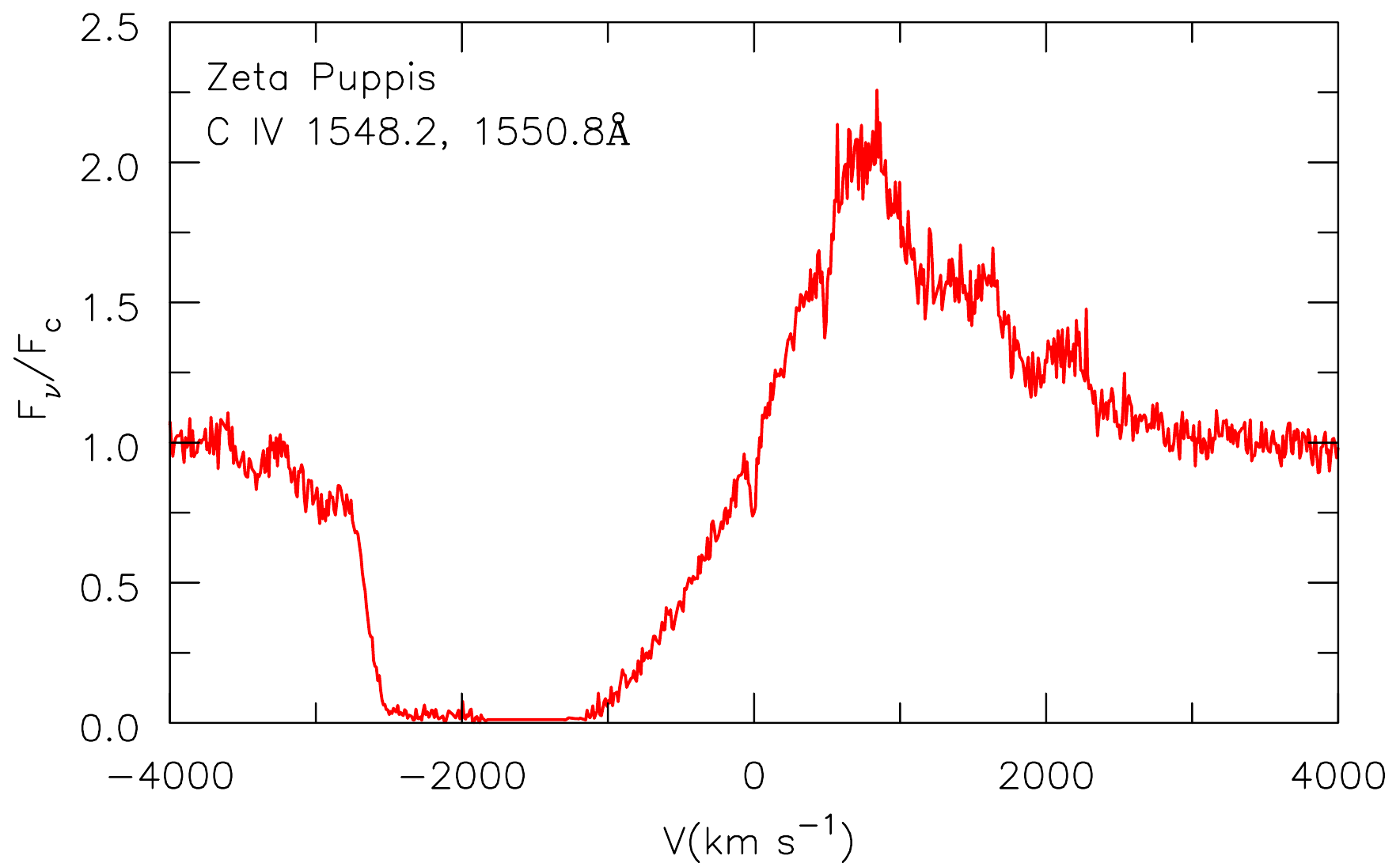}
\caption{The \ion{C}{4} $\lambda\lambda$1548.2, 1550.8 P~Cygni profile in the O4 I(n)fp star Zeta Puppis.
The x-axis was computed using $v=c(\lambda/\lambda\tsub{o}-1.0)$ where $\lambda\tsub{o}=1548.187$. The spectrum has been normalized by the continuum spectrum ($F_c$) -- i.e., a smooth curve drawn through spectral
regions showing no evidence for bound-bound absorption or emission.
Strong blue shifted absorption is seen, indicating an outflowing stellar wind with a terminal velocity (\Vinf)
in excess of 2600\,\kms. The redshifted emission primarily arises from continuum photons that were
emitted in other directions, absorbed by  \ion{C}{4}, and subsequently scattered to the observer.
The two narrow absorptions near 0 and 500\,\kms\ are due to absorption by \ion{C}{4}\ in the interstellar medium.}
\label{fig_pcyg}
\end{figure}

With the discovery of X-ray emission from O stars it was realized that X-rays could explain the presence
of both \ovires\  and \nvres\ P~Cygni profiles\footnote{A  P~Cygni  profile is formed when continuum radiation is absorbed and scattered by outflowing material.  Outflowing gas along the line of sight absorbs continuum radiation and scatters it out of the line of sight, producing blue-shifted absorption. Radiation absorbed in other directions can be scattered into the line of sight, and for a spherically symmetric expanding gas, , the combination with the blue-shifted absorption will give rise to red-shifted emission.} in the UV spectra of O stars. An
example P~Cygni profile is shown in Figure~\ref{fig_pcyg}. Since O stars typically have effective temperatures of $< \sim40,000$\,K, the photospheric radiation field cannot produce sufficient \ion{O}{6}\ to explain the observed \ion{O}{6}\ profile. However X-rays, through Auger ionization, can produce sufficient \ion{O}{6} \citep{CO79_auger}. In the case of \ion{O}{6}, the crucial reaction is:
\blankhalf
\indent
\ion{O}{4}(1s$^2$ 2s$^2$ 2p)+ X-ray $\rightarrow$ \ion{O}{5}(1s 2s$^2$ 2p) $+$ e$^-$  $\rightarrow$ \ion{O}{6}(1s$^2$\,2s) + 2e$^-$ 
\blankhalf
Typically two electrons are ejected (i.e, one by interaction with the photon, and one by the Auger process) in Auger ionization for CNO elements, but for heavier elements more than two electrons can be ejected \cite[e.g.,][]{Wei74_Xray_cross}. In \cmfgen\ we assume all inner-shell ionizations only eject two electrons, and the intermediate states are omitted.\footnote{The ejecta of Type Ia SNe are composed
primarily of intermediate mass elements (Ca, Si) and iron group (Fe, Ni, Co) elements. In such ejecta
we may need to  treat Auger ionization more rigorously since it could potentially affect the ionization state of
the gas and the thermalization of non-thermal electrons. The non-thermal electrons are initially
produced via Compton scattering of gamma-ray photons produced from decay of radioactive $^{56}$Ni and
$^{56}$Co. In this case inner shell ionization will most likely occur via non-thermal electrons. However the subsequent Auger ionization and fluorescence is independent of how the K-shell hole was created.} Many studies have shown that inner shell ionization
of X-rays can successfully explain the presence of \ion{O}{6}\ and \ion{N}{5}\ in O stars \citep[e.g.,][]{PHL01_WMbasic,Zsargo08_zpup}. Auger ionization complicates the kinetic equations since more than
two ionization stages are directly coupled.

\section{Recombination}
\label{Sect_recom}

The recombination rate is given by
\begin{equation}
               \left(dn_l\over dt\right)_{\scriptsize \rm RR}=  n\tsub{K}\left(n_l\over n\tsub{K}\right)^* \int_{\nu_o}^\infty \left({4\pi \over h\nu}\right) \sigma_\nu \left({2h\nu^3\over c^2}+J_\nu  \right) \exp(-h\nu/kT) d\nu
\end{equation}

\noindent
 \citep[e.g.,][]{Mih78_book} where the subscript K refers to the recombining ion, and
 the LTE population is computed using the actual electron density. The quantity $\left(n_l/ n\tsub{K}\right)^*$ (for a given level) is only a function of the electron density and temperature. When the gas is in LTE, and when $J_\nu=B_\nu$, the photoionization and recombination rates (absolute values) are
identical. 

In my work I treat recombination as the reverse process of photoionization, and hence in \cmfgen\ recombination rates are computed using the  photoionization cross-sections.  As noted earlier, rates are evaluated using numerical quadrature, and identical weights are used for both the forward and reverse process. At high densities it is desirable to treat both processes identically since small differences can cause erroneous populations to be determined when solving the kinetic equations.
At depth, where LTE conditions apply, it is important that they identically cancel.
Generally the weights are evaluated using the trapezoidal rule -- more accurate quadrature schemes are generally not feasible because of the complex frequency dependence (and depth dependence) of the radiation field, and because the same quadrature scheme must be used to compute the rates for both  photoionization and recombination. Care must be taken near bound-free edges, since the integrand in the recombination 
rate can vary rapidly with frequency -- especially true for highly ionized states at low temperatures
since the recombination rate at frequency $\nu$ scales as  $\exp[-h(\nu-\nu_o)/kT]$.


For low densities, such as those found in \ion{H}{2} regions, planetary nebulae, and many collisionally ionized plasmas, recombination rates are often evaluated separately, and treated as a distinct process. At ``low" densities most transitions are optically thin, and recombination into high states simply cascade into the ground state and metastable levels.  In a \ion{H}{2} region, for example, the ionization of H is maintained through photoionizations from the ground state, and photoionizations from excited states can be ignored. However transitions to the ground state can be optically thick. Consequently two limiting cases are considered when computing H line strengths -- Case A, in which all transitions are assumed to be optically thin, and Case B, in which only the Lyman transitions are optically thick \cite[e.g.,][]{2006agna.book.....O}.  Under the
optically thick assumption the rate of decays in a transition is assumed to be exactly balanced by
the rate of radiative excitations in the transition. 

\subsection{Direct radiative recombination}
\label{Sect_dir_recom}

This process is taken to refer to simple recombination processes such as

\blankhalf
\indent
H$^+$ + \emin  $ \rightarrow$ H(n$l$) + $h\nu $
\noindent
and
\blankhalf
\indent
\ion{C}{4}(2s $^2$S)  + \emin $\rightarrow$ \ion{C}{3}(2s nl) + $h\nu $ \,\,. 
\blankhalf
\noindent
In the above n and $l$ refer to the principal and angular momentum quantum numbers of the electron.

\blankhalf
The cross-sections for the processes are ``smooth" and easily integrated. However,
state-of-the-art photoionization cross-sections, such as those available through
\href{http://cdsweb.u-strasbg.fr/topbase/topbase.html}{TOPbase} and \href{https://norad.astronomy.osu.edu}{NORAD}, often include indirect photoionization routes, or resonances. When discussing the reverse process it is often convenient to split the  resonances into two classes -- low temperature dielectronic recombination (LTDR) resonances, and high temperature dielectronic recombination (HTDR) resonances. 

\subsection{Low temperature dielectronic recombination (LTDR)}
\label{Sect_LTDR}

LTDR  refers to recombination through double-excited states that lie close, but above, the ionization energy of the ion ground state \citep{NS83_LTDR,NS84_CNO_LTDR}.  It is treated separately from HTDR since it is very species/ionization state specific --  the energy levels involved in LTDR need to be known accurately since the distance of these states from the ionization limit is a crucial factor in determining the recombination rate.
Dielectronic recombination is the inverse process of autoionization.

\begin{figure}[htbp]
\centering
\includegraphics[width=7.5 cm]{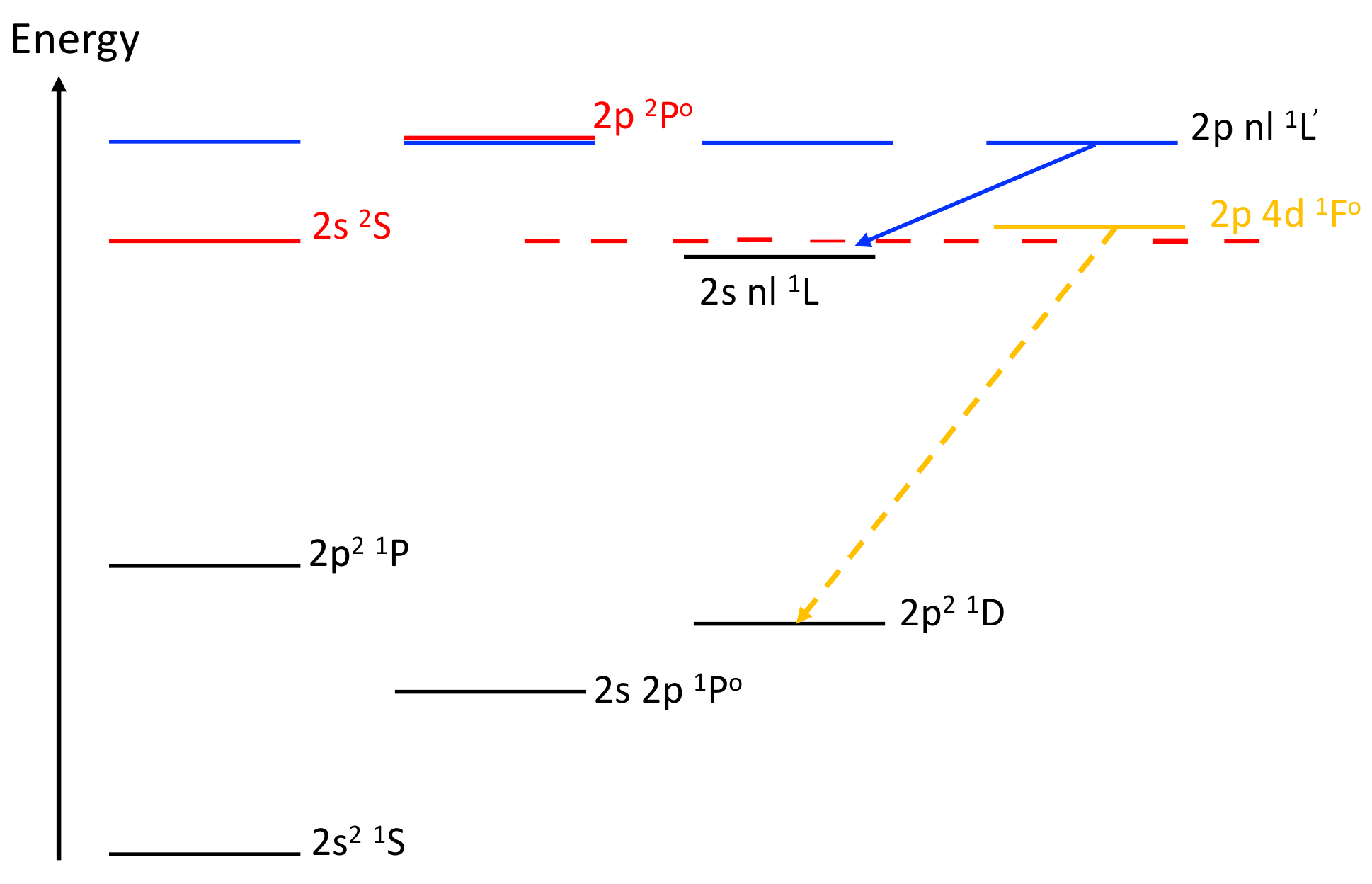}
\caption{Simplified pseudo-Grotrian diagram for \ion{C}{3} and  \ion{C}{4} to help
illustrate LTDR and HTDR.  Five bound  \ion{C}{3} levels are shown in black and two \ion{C}{4}\ levels
are shown in red. Example HTDR autoionizing levels (e.g., 2p 50p, 2p 50d, etc) that
converge on the \ion{C}{4} 2p $^2$P\tsup{o} state are shown in blue.  One of the most important
 LTDR  autoionizing levels is shown in orange -- it  has been moved up slightly in energy to separate
 it more clearly from the \ion{C}{4} ground state -- the horizontal dashed red line is used to indicate the energy of the \ion{C}{4} ground state . The blue line shows a HTDR transition, which has a wavelength approximately equal to that of the \ion{C}{4}\ resonance transitions (1548, 1551\,\AA), while the dashed orange line shows a LTDR transition
(at $\sim 412\,$\AA). Triplet levels also experience both LTDR and HTDR.}
\label{fig_ciii_grot}
\end{figure}

We can understand the LTDR process as follows by using an example. Consider the \ion{C}{3}\ 2p\,4f $^1$D state (Figure~\ref{fig_ciii_grot}) which can autoionize (the inverse process to dielectronic recombination) to give \ion{C}{4} and a free electron. For this state the autoionization rate coefficient is large ($> 10^{13}$\,s$^{-1}$).\footnote{From {\sc autostructure} calculations made by a collaboration of researchers at Auburn University, Rollins College, the University of Strathclyde, and other universities. Tables produced  by N.\ R.\ Badnell and are available
at \href{http://amdpp.phys.strath.ac.uk/tamoc/DATA/} {Atomic Data from AUTOSTRUCTURE}.} However the level can also undergo a ``stabilizing transition", leading to a recombination, with the most important stabilizing transition being 
\blankhalf\indent
2p\,4d \odd{1}{F}\ $\rightarrow$ 2p$^2$ $^1$D + $h\nu$\,\,.
\blankhalf 
The Einstein A coefficient for this process ($A_{ul}$) is $\sim 6.2 \times 10^9\,{\rm s}^{-1}$ (P.\ J.\ Storey, private communication),  much lower than the autoionization probability. Consequently the 2p\,4f\,\odd{1}{F}\ state will be in LTE with respect to \ion{C}{4}, and hence the LTDR recombination rate for this single transition is   $n_u^* A_{ul}$ where  $n_u^*$ (in cgs units) is given by
\begin{equation}
n_u^*= {2.07 \times 10^{-16} \over T^{3/2}} {g_u \over g_{\scriptstyle\rm C IV}} N_e N_{\scriptstyle\rm C IV} \exp{(-\psi_l/kT)}
\end{equation}

\noindent
\citep[e.g.,][]{Mih78_book}. In the above formula $g_u$ is the statistical weight for the 2p\,4d
\odd{1}{F}\  state, $g_{\sc C IV}$ is the statistical weight of the \ion{C}{4}\ ground state (2s\,$^2$S), $N_{\sc e}$ is the electron density, 
$N_{\sc C IV}$ is the ground state population of \ion{C}{4}, and $\psi_l$ is the energy of the  2p\,4d\odd{1}{F}\  state above
the ground state of \ion{C}{4}.  At $10^4\,K$ that single transition leads to a LTDR recombination
coefficient (defined as the (LTDR rate)$/N_e/N_{\sc C IV}$) of  $3.1\times 10^{-12}$\,cm$^3$\,s$^{-1}$ (P. J. Storey, private communication) which is essentially identical to the direct recombination rate of $3.2\times 10^{-12}$\,cm$^3$\,s$^{-1}$  \citep{1973A&A....25..137A}.

Thus we see the following:
\begin{enumerate}
\item 
       The LTDR rate is very sensitive to $\psi$ when $\psi_l/kT$ is of order unity, or larger.
\item
     When $\psi_l/kT << 1$, the LTDR recombination rate scales as $T^{-3/2}$, and thus
                increases more quickly with decreasing temperature than the radiative recombination rate, which typically scales as $T^{-\alpha}$ with $\alpha \sim 0.7$. \citep[see, e.g.,][]{1973A&A....25..137A}.
\item
      The LTDR process will be most important for those states with a large Einstein A coefficient, and for those states lying closest to, but above, the ion ground state.
\item
      The process is very dependent on the details of the atomic structure. In the above case, the energy of the 2p\,4d \odd{1}{F}\ state is crucial for determining the LTDR rate. As the LTDR autoionizing states lie well above the \ion{C}{3}\ ground state, and can have large energy widths, the energies of the states are not necessarily known. Theoretical calculations can provide estimates, but will have difficulties for states that lie "very close" to the ionization limit since a small error in the energy level can make a big difference in the recombination rate, particularly at low temperatures. 
 \end{enumerate}

The LTDR rate can exceed the direct recombination rate, and in many cases plays a crucial role
in determining nLTE level populations, and observed line strengths
\citep[e.g.,][]{NS83_LTDR,NS84_CNO_LTDR,2013ADNDT..99..633S}.

The LTDR process is complicated by states that are forbidden to autoionize in LS coupling, such
as the 2p\,\,4d\,\,$^3$D$^{\sc o}$  state in \ion{C}{3} or the quartet states in \ion{C}{2}. In such cases the populations of these levels must be determined by solving the rate equations. These levels will be collisionally and radiatively coupled to states that can autoionize and because of departures from LS coupling, they  can also have non-zero autoionization rates which are  larger than the radiative decay routes from the state. Thus, these levels can be an important additional recombination channel.

In \cmfgen\ we handle the quartet states in \ion{C}{2} as part of our atomic models
while the doublet autoionizing states are assumed to be in LTE with respect to
the ground state of \ion{C}{3} and are not directly
treated. Recombination through the  quartet states is  treated via the line transitions
connecting them to lower levels, while transitions for the autoionizing states are treated via
the photoionization cross-sections. Generally we assume the states within a term are
populated according to their statistical weights, although this  will not be valid
for some levels since the autoionizing rates can depend strongly on their total angular momentum. For example,  the autoionizing probabilities for the \ion{C}{2} 2s 2p(\odd{3}{P}) 4s\,\odd{4}{P}\  j=1/2, 3/2, and 5/2 states are $5.3\times 10^9$, $1.3 \times 10^{10}$, and $< 3\,{\rm s}^{-1}$. These were obtained from the full-width at half maximum tabulated by \cite{2013ADNDT..99..633S}. One issue, potentially important at high densities, is that we do not have accurate collisional cross-sections for the states not permitted to autoionize in LS coupling.

For \ion{C}{3} we typically assume, following \cite{NS84_CNO_LTDR}, that all low lying states can autoionize.  LTDR is easily taken into account via the photoionization cross-sections, however the assumption is only necessarily valid for those states in which the autoionization rates (greatly) exceed other processes populating/depopulating the autoionizing state.

A potential problem in nLTE calculations is that, due to difficulties of current atomic codes to compute accurate energies, the resonances in the photoionization cross-sections are offset from their true positions. Such offsets are probably unimportant when computing the Rosseland mean opacity, but can be important for spectral studies. First, an inaccurate energy will influence the location of observable resonances in stellar spectra. Second, it can have an effect at ``low" temperatures due to the scaling of the LTDR rate with temperature ($\exp(-\psi/kT)/T^{3/2}$). Third,  complicated nLTE effects could arise. For example, a wrong
resonance wavelength can potentially cause issues if a strong resonance coincides (or now doesn't coincide) with another bound-bound transition (since a strong resonance can affect the radiation field in the transition and vice versa).

The direct inclusion of resonances in photoionization cross-sections also has  other potential issues. First,
the resonances vary much more rapidly than the background cross-section, and hence a very fine frequency grid needs to be used --- this is particularly true for narrow resonances. For computational expedience,
we typically sample the continuum cross sections  in \cmfgen\ every 500\,\kms\ (but finer near level edges). To avoid aliasing\footnote{A signal processing term that refers to the distortion of data due to sampling which is too coarse. In the present case a narrow but strong resonance could be missed in the photoionization cross-section when the frequency sampling is too coarse. Alternatively, its influence could be artificially enhanced if it is not fully resolved.}
we smooth the cross-sections. In early versions of the atomic data the cross-sections were
smoothed to a resolution of 3000\,\kms, but in new data sets we no longer store the smoothed cross-sections.
Instead, newer cross-sections can be smoothed to the desired resolution, set by a control parameter, when they are read in. 

Second, a narrow resonance can mean that  the autoionization lifetime of the upper level may be  comparable, or even larger, than radiative transitions from the same level. As a consequence the upper level may not be in LTE with respect to the ion, and hence the photoionization cross-section should not be used to  compute the recombination rate. When identified, such a resonance should be clipped out, and the upper levels treated as a bound state.

Third, photoionization cross-sections are usually computed in LS coupling. This means, for
example, that the multiplet structure of the resonances is not treated -- a problem more crucial
when the resonances are ``narrow''.

\subsection{High temperature dielectronic recombination (HTDR)} 
\label{Sect_HTDR}

HTDR involves high Rydberg states \citep{1965AnAp...28..774B,1965ApJ...141.1588B}, and it is very difficult to treat accurately in stellar atmosphere codes. The easiest way to visualize HTDR is to discuss a specific example.


Consider for example \ion{C}{3}\ whose ground state is 2s$^2$ $^1$S where we have omitted the complete inner shell for simplicity. The states contributing to HTDR are the
Rydberg states of the form 2p~n$l$ that converge on the \ion{C}{4} state 2p \tsup{2}P\tsup{o}
(Figure \ref{fig_ciii_grot}).
Such states can autoionize to give \ion{C}{4} 2s $^2$S, or the 2p electron can
radiatively decay giving rise to \ion{C}{3} 2s\,n$l$. At ``low" densities the n$l$ electron
will decay to a lower level, producing a ``real" recombination. Since the autoionization
rates decay slowly with n, and since  the 2p $\rightarrow$ 2s transition probability is approximately constant, high n values (e.g., up to n$=100$)  determine the net recombination rates.
Such levels are not typically included in nLTE calculations. The process is
further complicated because the autoionizing rates strongly depend on the
angular momentum -- low ``$l$" states have much higher autoionization 
probabilities than do higher angular-momentum states 
\citep{Bur64_HTDR,1965AnAp...28..774B}.

At low densities the HTDR recombination will scale roughly as $\exp(-E/kT)/T^{3/2}$ where
$E$ is the energy of the 2s-2p transition in \ion{C}{4}. Since the energy of
the 2s-2p transition is well known, and since the energy of the high Rydberg states 
is easily approximated,  the accuracy of  the energy levels is not a crucial factor
determining HTDR rates. For some species, several Rydberg series may
contribute to the HTDR rate, yielding a more complicated temperature dependence
than the simple expression provided above.

\begin{figure}[htbp]
\centering
\includegraphics[width=7.5 cm]{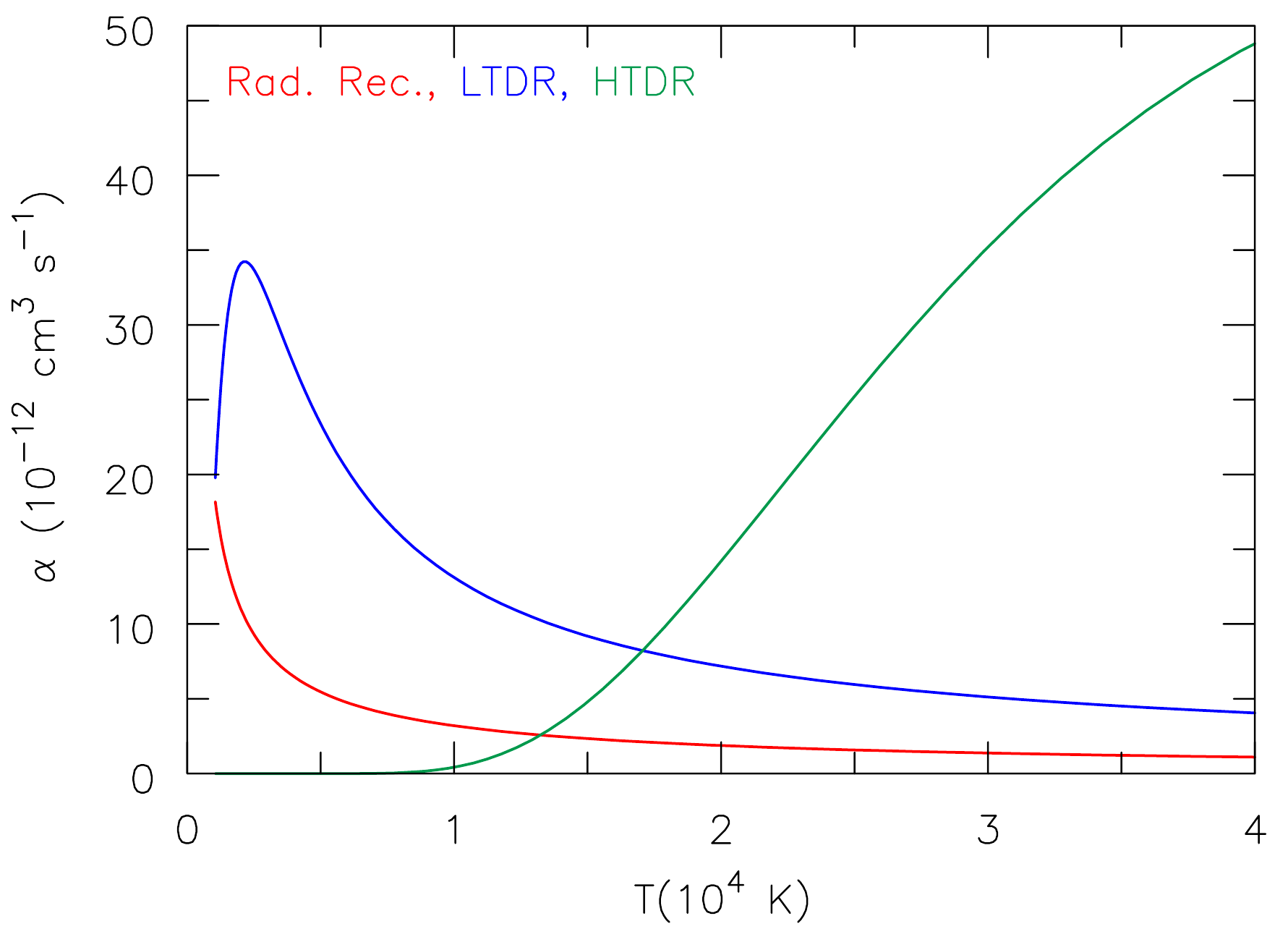}
\caption{Comparison of the recombination coefficients for normal radiative
recombination, for LTDR, and for HTDR for recombination from C$^{3+}$ to C$^{2+}$. Only at low temperatures ($\lesssim 1000\,$K) does
radiative recombination dominate. Above this temperature LTDR dominates  until a
temperature of $\sim 17,000\,$K, at which time HTDR becomes the dominant recombination mechanism.}
\label{fig_ciii_alpha}
\end{figure}

In Figure~\ref{fig_ciii_alpha} the different recombination coefficients for \ion{C}{3}\ are plotted. The radiative recombination rate, and the HTDR rate are from \cite{1973A&A....25..137A}, while the LTDR rate is from \cite{NS83_LTDR}. The crucial importance of dielectronic recombination for \ion{C}{3} is clearly demonstrated by this figure.

\section{Suppression of dielectronic recombination}
\label{Sect_suppress}

\subsection{Collisional processes}
\label{Sect_sup_col}

The classic  HTDR formula only applies at low densities. As the density rises, collisional ionization by electrons can significantly suppress HTDR. This has been explicitly considered for HTDR of C$^{3+}$ to C$^{2+}$ 
(i.e., \ion{C}{4} to \ion{C}{3}) by \cite{1975ApJ...195..285D,1993ApJ...407L..91B}.  \cite{1993ApJ...407L..91B} found suppression factors of $\sim 0.7$ at $10^4$,  $0.4$ at $10^6$, $0.2$ at $10^8$ and $0.1$ at $10^{10}$\,cm$^{-3}$ (these data were read from figure 1 in  \cite{NGK13_HTDR}). The reduction in the HTDR rate arises because the stabilizing transition (2p-2s in the case of \ion{C}{3}) from the autoionizing states leaves the electron in a high n$l$ state. As the electrons cascade down to lower
n$l$ states, they can be collisionally ionized by electrons.



\subsection{The importance of the Radiation Field}
\label{Sect_sup_rad}

The radiation field is typically regarded as unimportant in the dielectronic process, but in
stellar atmospheres and winds, the process could effectively suppress recombination,
as illustrated below. In principle, there are two suppression routes. The radiation
field can directly suppress the stabilizing transition, or the radiation field can directly
ionize an electron out of the high n$l$ state which the stabilizing transitions have left
the ion in. The latter is similar to collisional suppression, except it is the
radiation field, rather than collisions with electrons, which is reionizing the atom/ion.

For simplicity we treat the resonance as a line transition between two  bound states, with the upper level being the autoionizing level.
The net recombination rate will be given by
\begin{equation}
NR= n_u A_{ul}\left( 1- \Jbar/S_{lu} \right)
\end{equation}
where $\Jbar$ is the mean intensity in the line and is given by
\begin{equation}  \label{Eqn_jbar}
 \Jbar = \int_0^\infty \phi(\nu) \,J_\nu \,d\nu\,\,,
\end{equation}
$\phi(\nu)$ is the line absorption/emission profile (which, in general,  is determined by the finite lifetimes
of the levels involved in the transitions, thermal motions of the atoms, and the interaction of
the radiating atom/ion with its neighbors, \citep[see][chapter 9]{Mih78_book} and $S_{lu}$ is the line source function given by
\begin{equation}
 S_{lu} =  {2h\nu \over c^2} \left( {n_u/g_u \over n_l/g_l -  n_u/g_u}\right)
\end{equation}

\noindent
In LTE $n_u=(g_u/g_l) n_l \exp(-h\nu/kT)$ and hence $S_{lu}$ simplifies to the Planck function.
As readily apparent the net rate does not directly  depend on the  optical depth -- such a dependence only occurs  indirectly through the dependence of  \Jbar\ on the optical depth.

In nebula conditions $\Jbar/S_{lu}$  is typically $<< 1$ (since the nebula
is very distant from the star the radiation field is greatly diluted), and the contribution to the recombination rate by this single transition is simply $n_u A_{ul}$. When the rates are summed over all resonances you recover the LTDR/HDTR recombination rate. However such a rate is typically an upper limit since the radiation field can reduce this rate.

\begin{figure}[htbp]
\centering
\includegraphics[width=16.0 cm]{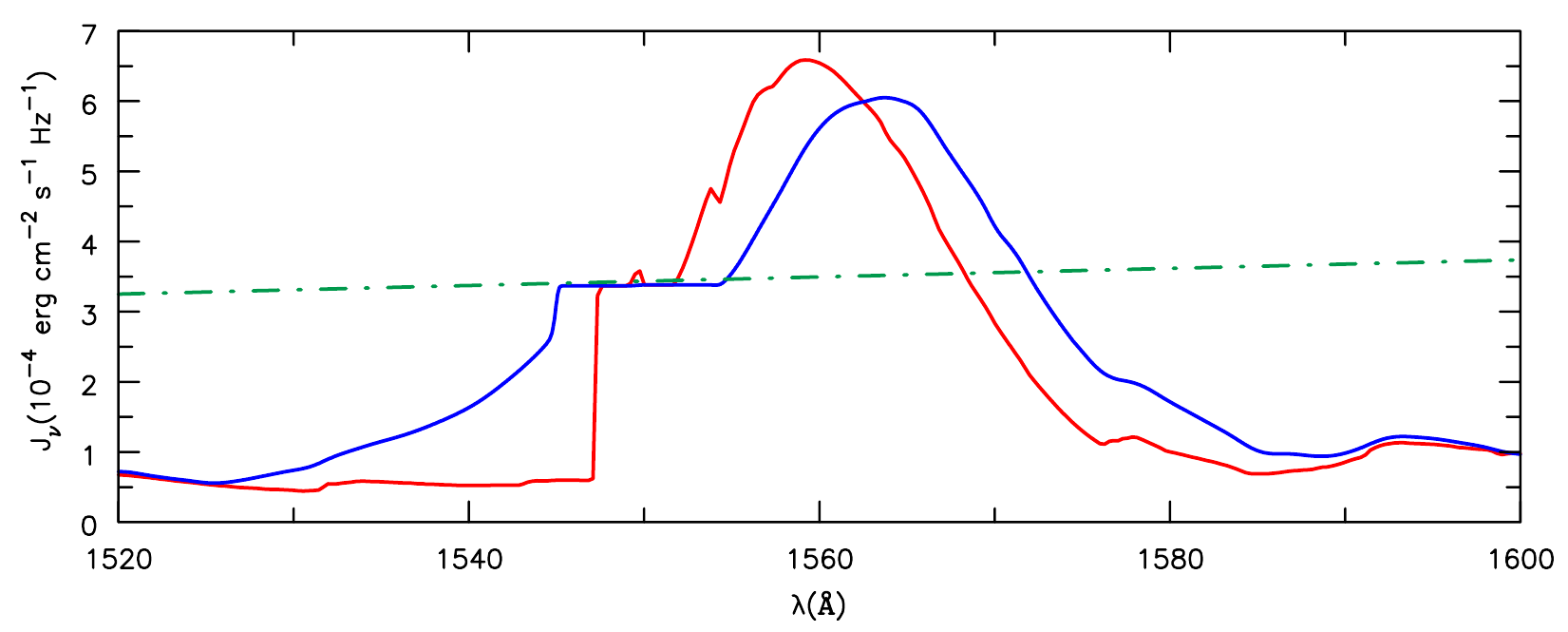}
\caption{Comparison of the mean intensity (in the comoving frame, red curve) and the blackbody mean intensity (green-dashed curve) at a temperature of $1.6\times 10^4\,$K and  a density of $\sim10^{11}$ electrons cm$^{-3}$ -- roughly 50\% of the emission in \ion{C}{3}\ $\lambda 2297$ originates above that density. At this density the \ion{C}{4}\ 2s and 2p states are collisionally coupled, and as a result the radiation field near the wavelengths of the \ion{C}{4}\ transitions at 1548, 1551\,\AA\ is also close to that of a blackbody. The
blue curve is similar to the red curve except that we used a Voigt profile for the line absorption/emission profile
(see Equation~\ref{Eqn_jbar}), rather than the simple Doppler profile \citep[see][]{Mih78_book} which is generally used when computing the atmospheric structure and level populations. The use of a Voigt profile is crucial for explaining the observed profile of the \ion{C}{4}\ resonance doublet at 1548, 1551\,\AA\ in WC stars \citep{H89_WC}.
}
\label{fig_j_bb_comp}
\end{figure}

At depth in a stellar atmosphere $\Jbar \equiv S_{lu} \equiv B_\nu$, and thus the net LTDR and
HTDR rates are identically zero -- that is, every downward transition in the stabilizing
transition is balanced by an upward transition. However above the atmosphere the temperature of the radiation field and the electrons are not the same. Typically  $\Jbar$  will fall below  $S_{lu}$, however
in a wind $\Jbar$ can be greater than $S_{lu}$ in some transitions.  In Figure~\ref{fig_j_bb_comp}
we show the mean intensity (in the comoving frame) and the blackbody mean intensity
at a temperature  of $1.6\times 10^4\,$K and  a density of $\sim10^{11}$ electrons cm$^{-3}$ -- roughly 50\% of the emission in the line referred to as \ion{C}{3}\ $\lambda 2297$ originates above that density.  From that figure we see that the radiation field at the wavelength of the \ion{C}{4} resonance transition, and at/near the stabilizing transition, is close to a blackbody at the local electron temperature. Thus the radiation can act to suppress HTDR.

\begin{figure}[htbp]
\centering
\includegraphics[width=16.0 cm]{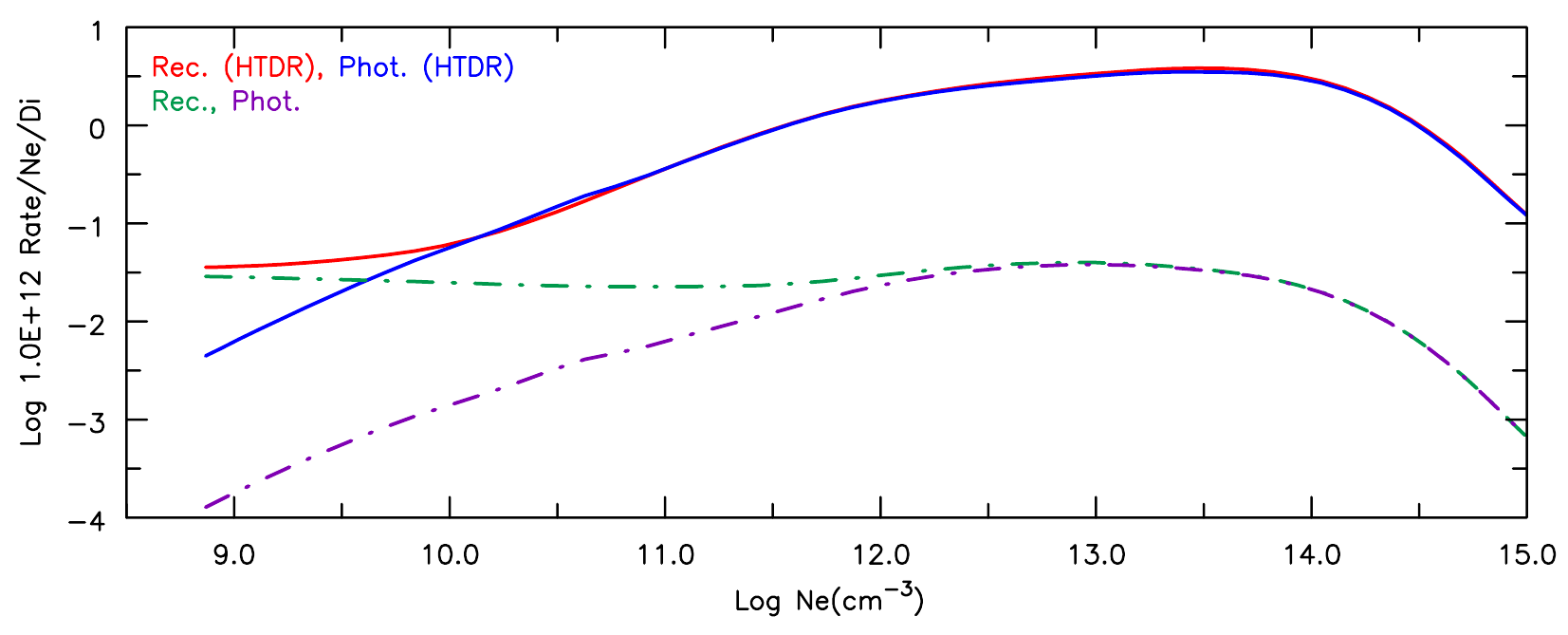}
\caption{Illustration of the  photoionization and recombination rates to/from a combined level ($n=26-30$;  singlet levels only) in \ion{C}{3}. The solid curves show the rates with HTDR included, while the dashed curves show the rates without HTDR included. At most densities the recombination and photoionization rates agree
closely in the model with HTDR. In such regions the influence of HTDR on the ionization structure
is suppressed. Towards the outer boundary the HTDR recombination rate converges to that of a
model without HTDR due to the decrease in electron temperature.}
\label{fig_HTDR}
\end{figure}

 In Figure~\ref{fig_HTDR} we show the recombination and photoionization rates for $n=26$ through 30 singlet states of \ion{C}{3} (treated as a single level) for a test calculation in which we included HTDR transitions for levels up to $n=30$, and with no suppression of the recombination rate with the angular orbital quantum number. The resulting model spectrum is almost identical to the spectrum computed without HTDR -- a consequence of the low temperatures in the \ion{C}{3} line-formation region and the suppression of HTDR via the radiation field in the stabilizing transition. This result is model dependent -- in practice the importance of HTDR needs to be examined on a case-by-case basis.

 
The influence of the radiation field is not an issue if the resonance is included as part of the photoionization cross-section as the influence of the radiation field is automatically taken into account. It is also not an issue for
``autoionizing" states treated as bound levels, since the radiation field is again taken into
account. However, it is  a potential issue if the LTDR or HTDR rate is included as a
separate process, and the inverse process is not included.

\section{Astrophysical examples}
\label{Sect_examp}

Below we discuss some examples of where  photoionization data is crucial. It
is unrealistic to discuss all cases, since photoionization data is crucial in any
photoionized plasma, and is crucial for nLTE analyses.   For some plasmas, 
in which collisional processes dominate, photoionization is less important, but the inverse
process (recombination) is still critical.

\subsection{Recombination processes}
\label{Sect_ex_recom}

The importance of photoionization/recombination processes depends critically on the application.
Here we discuss photoionized plasmas,  and gradually work our way up in density.

In ionized nebulae and \ion{H}{2}\ regions H and He lines are produced by recombination, while the ionization is typically maintained via ionizations from the ground state. The strongest metal lines (i.e., those not due
to H and He), such as \ion{O}{3} and  \ion{N}{2}, typically arise via collisional excitation from the ground state.
Metal recombination lines are much weaker, simply because the metal abundances are typically a
factor of $10^3$ (or more) lower than that of H.  Some other lines are produced by line fluorescence, where the radiation field in a bound-bound transition in one species gives rise to line emission in another species (the
Bowen mechanism). For example, some \ion{O}{3}\ lines in planetary nebulae (gaseous
nebulae surrounding stars with effective temperatures in excess of 30,000\,K) are produced by the chance overlap of an \ion{O}{3}\  line with \ion{He}{2} Ly$\alpha$ \citep{Bow34_fluro,2006agna.book.....O}.
With high signal-to-noise spectra metal recombination lines are seen \citep[e.g.,][]{2013MNRAS.429.2791F,2017PASP..129h2001P}, and for most lines their strength is effectively set by optically thin recombination (radiative and dielectronic) theory. 

As the density increases, processes become increasingly complex. More transitions become
optically thick, affecting the cascade process and hence line strengths. Collisional coupling
between the levels also becomes more important. If the radiation field is not too diluted,
photoionization from excited states also becomes increasingly important. For example,
in O and WR stars, the ionization of He$^+$ to He$^{++}$ occurs predominantly from the n$=2$
state whose population is maintained via the intense radiation field in Ly$\alpha$ \citep{Hil87_B}. Similarly, the ionization of \ion{C}{4}\ is maintained  from the $n =3$ levels. 

As the density further increases, lines become thicker, and photoionization/recombination
can become more important since continuous processes at many wavelengths remain optically thin.
In hot stars the departure coefficients ($=n/n^*$) for H and \ion{He}{2}\ levels typically rise above the photosphere.  Bound-bound transitions are optically thick, preventing cascades. The radiation is diluted, hence recombinations into a level typically exceed photoionizations from that level. Eventually, however, photon escape in lines becomes important, and the departure coefficients decrease. However one must be careful with
generalizations -- at some wavelengths (particularly in the Wien regime of the blackbody curve) the rapid fall of the electron temperature with height above the photosphere means that the energy density in the radiation
field may initially exceed the radiation energy density predicted by the blackbody formula for the local electron temperature. 

Bound-bound processes are crucial for determining line strengths. However, it is
ultimately photoionization and recombination that determines the ionization state.
In some cases, charge exchange processes are crucial \citep{1971ApJ...166...59F,1973MNRAS.164..111W}. Particularly important are charge exchange process of  neutral H with, for example,
Fe$^{2+}$ and O$^{+}$. The reaction \\
$${\rm O}^{+} + {\rm H} \rightleftharpoons  {\rm O}  + {\rm H}^{+}$$
is resonant, has a total rate coefficient of order $10^{-9}$\,cm$^3$\,s$^{-1}$ \citep{1999A&AS..140..225S}, and is crucial for determining the {\rm O}$^{+}$/{\rm O} ratio in regions where the neutral hydrogen
fraction exceeds roughly $10^{-3}$.

\subsection{The Sun}
\label{Sect_sun}

For the Sun, our nearest star, we can determine its structure in two ways. First, we can use the
observed Solar parameters (M, L, \Rstar, abundances) to construct a theoretical model of the Sun.
Second, we can use helioseismology observations to constrain the internal structure\footnote{
The Sun is simultaneously oscillating in thousands of different vibration modes.  The frequency
and strength of these modes depends on the internal structure of the Sun (e.g., the depth of
the convection, the sound speed)}. Unfortunately, the structure determined from
theoretical models and that determined from the helioseismology observations are inconsistent. They can be reconciled if the adopted opacities (for the relevant temperatures) are  too low. This could arise
 if the adopted O abundance is too low, or alternatively it could arise from inaccuracies
 in the opacities (i.e., inaccuracies in the photoionization cross-sections, oscillator strengths etc). The resolution of the problem is still unclear \citep{2021LRSP...18....2C,2019PhRvL.122w5001N,2015Natur.517...56B}.

\subsection{O, WR and LBV stars}
\label{Sect_OWR}

O stars are  the most luminous hydrogen-core burning stars known. They have masses in the
range 30 to $\sim 100$\,\Msun, and luminosities typically greater than $10^5$\,\Lsun. Due to nuclear processing  H is being converted to He in the core. At the same time most of the C and O initially present has been converted to N. Mass loss, and mixing, then operate to reveal this nuclear process at the stellar surface. During later evolution stages He is converted to C and O, and mass loss can also reveal this material at the stellar surface.

All massive stars are losing mass in a stellar wind.  In O stars, and their descendants
(e.g., LBVs,  WR stars)  the winds are driven by radiation pressure. Due to their high luminosities the stars are close to the Eddington limit\footnote{At the Eddington limit  the force arising from the scattering of radiation by free electrons matches the gravitational force.}. Consequently, it is relatively easy for radiation pressure acting through bound-bound transitions to drive material
off the surface of the star via a stellar wind.  Due to instabilities in the line driving, it is believed that the winds are highly clumped \cite[e.g.,][]{OCR88,1995A&A...299..523F,2018A&A...611A..17S}. Additional evidence for clumping comes from variability studies \cite[e.g.,][]{ELM98_zpup,LM08_var}, from the anomalously low
strength of some UV resonance transitions relative to the level of H$\alpha$ emission \cite[e.g.,][]{CHE02_FUSE,HLH03_AV83,BLH03_ngc340}, and the weakness
of electron scattering wings associated with strong emission lines
in P~Cygni stars and WR stars \footnote{The strength of most emission lines
in WR stars is proportional to the density squared. Thus a clumped wind can yield the same line
strengths for a lower mass-loss rate (i.e., for a lower average density).  On the other hand 
electron scattering line wings arise from Thomson scattering of line photons by free electrons and
hence scale with density.  Thus the strength of electron scattering wings relative to their neighboring
emission line can act as a global diagnostic of clumping. In WR stars the wings are offset to the
red from their originating transition because of the large outflow velocities.} \citep{Hil91_es,HM99_WC}.

The wind density in massive stars varies considerably. For main sequence O stars the winds
are relatively weak, and only affect a few spectral features. Their photosphere is geometrically thin
(i.e., $<<$ radius of the star), and, in principle, can be modeled using plane-parallel model atmospheres (i.e., the curvature of the star's atmosphere can be ignored), although the wind may still have an influence at some wavelengths. As the stars evolve, the wind density tends to arise and become increasingly important, and the use of a plane-parallel atmosphere is no longer valid.  Indeed, in WR stars, the wind is so dense that spectrum formation occurs in the  stellar wind and nLTE spherical models that treat the wind are essential.

\subsubsection{\ion{N}{3} and \ion{N}{4} lines in Of and WN stars}
\label{Sect_OfWN}

Of stars are evolved O stars that show emission in \ion{N}{3}\ and \ion{He}{2} $\lambda 4686$ \citep{1971ApJ...170..325C,1971ApJS...23..257W}. First computations of model atmospheres suggested that the \ion{N}{3}\ lines are driven into emission by LTDR \citep{MHC72_NIII}. However, more recent work that includes line blanketing (by lines of iron group elements) and winds reduces the
importance of dielectronic recombination, and continuum fluorescence acting through UV resonance
transitions plays a crucial role \citep{RPN11_NIII}.\footnote{In this process  a strong transition (typically in the UV) absorbs continuum photons, a process whose efficiency is enhanced by the velocity field which allows the UV transitions to intercept more continuum radiation. In many cases the absorbed photons will typically be re-emitted in the same transition. However, in some cases the upper levels have an alternate decay route -- decay via this transition can then lead to emission in this bound-bound transition.
This is also known as the Swings mechanism \citep{Swi48_Of}.} 

\begin{figure}[htbp]
\centering
\includegraphics[width=16 cm]{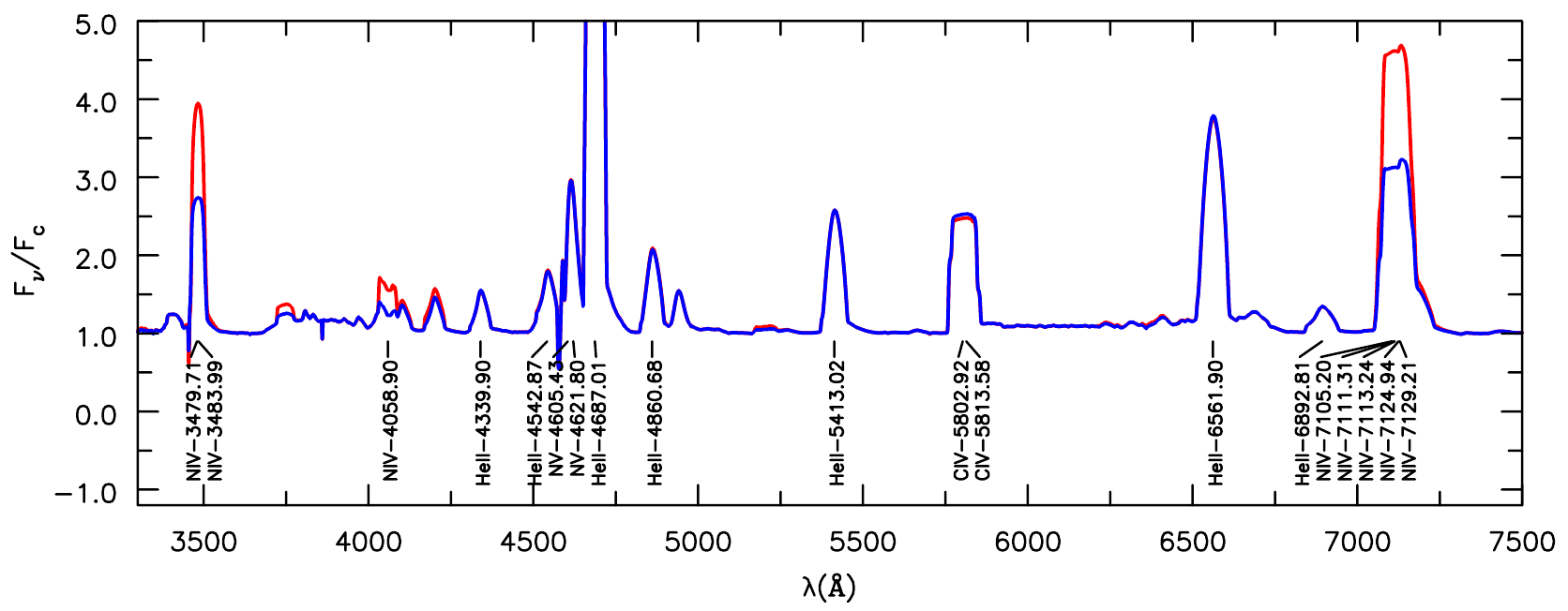}
\caption{Illustration of the influence of LTDR on the strength of several \ion{N}{4}\ emission lines.
 The model in red is the full calculation, while for the blue spectrum we omitted LTDR transitions
 in the calculation of the nLTE populations.  A \ion{N}{4}\ transition at 1718\,\AA\ is also influenced by
 LTDR. For these calculations we used smooth \ion{N}{4}\ photoionization cross-sections and the LTDR  data of \cite{NS83_LTDR,NS84_CNO_LTDR}. Calculations using the opacity photoionization cross-sections of \cite{TSB90_Be_seq}, that were obtained from \citep{Topbase93}, yield a spectrum very similar to that shown in red. The model has a luminosity of $3\times 10^5\,$\Lsun, a radius (at a Rosseland optical depth of  2/3) of 2.9\,\Rsun, an effective temperature of $\sim$78000\,K, a mass-loss rate of  $\Mdot=1.5\times 10^{-5}$\,\Msunyr, and a volume filling factor (which characterizes the  degree of clumping in the wind) of 0.1.}
\label{fig_niv_die}
\end{figure}

WN stars, which are a type of WR star, are evolved O stars which show abundances which have been influenced by the CNO nuclear burning cycle  -- H is depleted (in many it is absent), He  is enhanced, and much of the C and O has been converted  to N. In a WN star such as HD~50896, several \ion{N}{4}\ lines are seen. The formation of these lines is complex, but typically their strength is determined by a combination of dielectronic recombination and continuum fluorescence \citep{1988ApJ...327..822H}. While the models used by \citep{1988ApJ...327..822H} did not include iron group elements, more recent models with iron-group elements confirm the importance of LTDR for WN stars. In Figure~\ref{fig_niv_die} we illustrate the influence of LTDR on several  \ion{N}{4}\ emission lines for a model appropriate to an early-type WN star (such as HD~50896).
 
\subsubsection{Carbon in WC stars}
\label{Sect_C_WC}

WC stars are the evolved descendants of WN stars. Due to extensive mass loss, and nuclear processing in the interior of the star, their atmospheres are devoid of H, and are primarily composed of He, C and O (with similar mass fractions). Due to their dense stellar winds, and high C and O abundance, emission lines of He, C, and O dominate the spectrum. In the optical region, most of these arise from recombination, although optical depth effects greatly complicate line formation \citep{H89_WC}.


Below we discuss the spectrum of the WC4-type star, BAT99-9, which has recently been discussed by
\cite{2021MNRAS.503.2726H}.  In most ways its spectrum, and parameters, are typical of other WC4 stars in the Large Magellanic Cloud (LMC). However, it does differ in one important aspect -- it still exhibits one \ion{N}{5}\ and two \ion{N}{4}\ emission lines. Nitrogen is expected to disappear rapidly as a star transitions from WN to WC because N in the interior of the star is converted to $^{22}$Ne.

\begin{figure}[htbp]
\centering
\includegraphics[width=7.5 cm]{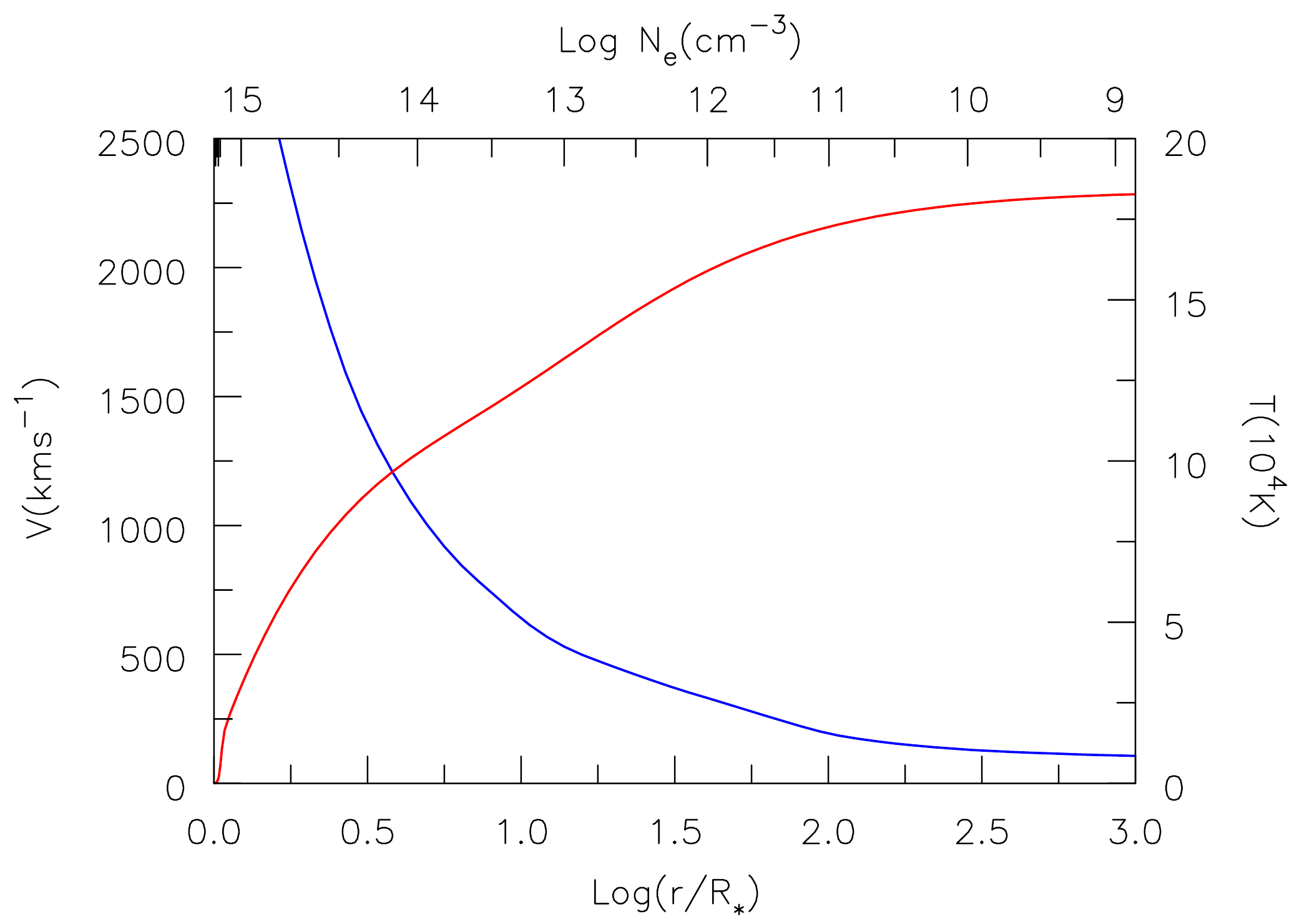}
\caption{Illustration of the electron temperature (blue) and velocity structure (red) in the  model for
the WC4-type star, BAT99-9. The model has $\log L/\Lsun=5.48$,  an effective temperature of 84,000\,K,
and a  mass-loss rate of $1.4 \times 10^{-5}\,\Msunyr$.}
\label{fig_vtemp}
\end{figure}

\begin{figure}[htbp]
\centering
\includegraphics[width=12 cm]{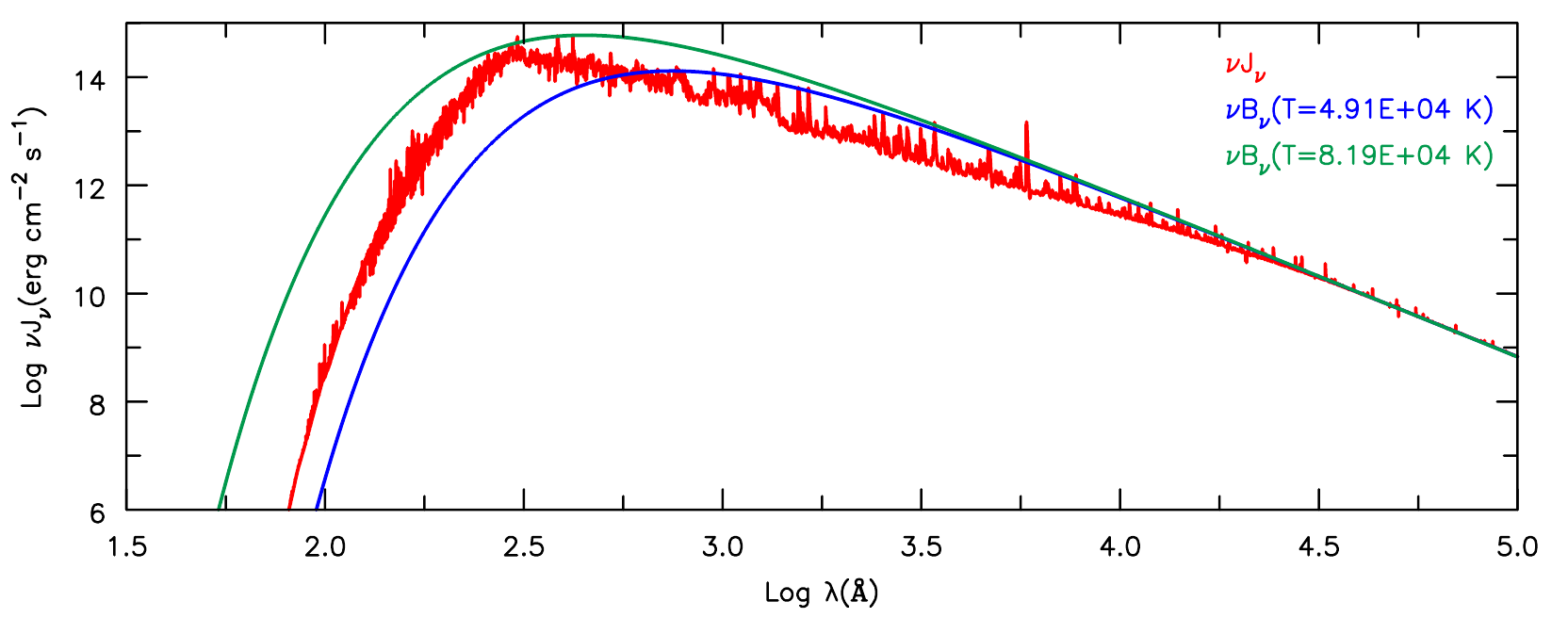}
\includegraphics[width=12 cm]{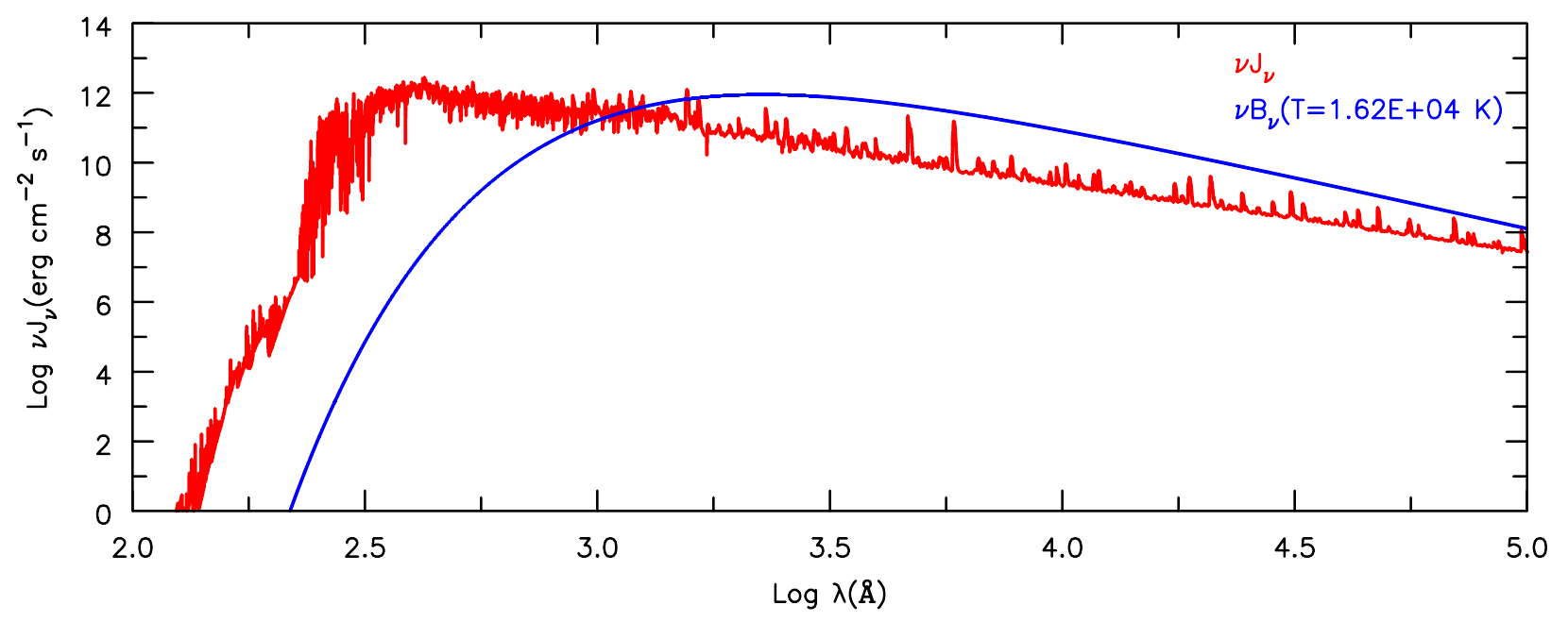}
\caption{An Illustration of the radiation field in a WC4 star at a depth where $V(r)=0.67 \times \Vinf$ 
$(N_e=1.7 \times 10^{13}\,{\rm cm}^{-3})$ (top panel)  and at a depth $V(r)=0.934 \times \Vinf$ $(N_e=1.2 \times 10^{11}\,{\rm cm}^{-3})$
(lower panel).   Also shown is the blackbody spectrum at the local electron temperatures ($T_e=4.90 \times 10^4\,$K and $T_e=1.61 \times 10^4\,$K).  In the top panel (where $V(r)=0.67 \times \Vinf$) we also illustrate the blackbody spectrum at the effective temperature ($T_e=8.19 \times 10^4\,$K)
that has been normalized to match the model spectrum at 5.0\,$\mu$m. As readily apparent, the local radiation field is very different from that defined by the blackbody at the local electron temperature, or that defined by the effective temperature. Consequently, and because of the low electron densities, the ionization state of the gas, and the level populations, are far from their LTE values. Due to the strong departures from LTE accurate atomic data is crucial for determining the state of the gas and hence for predicting the stellar spectrum.}
\label{fig_wc4_rad_field}
\end{figure}

The electron temperature structure and wind velocity of a model for the LMC WC4 star, BAT99-9, is shown 
in Figure~\ref{fig_vtemp}. The non-Planckian nature
of the radiation field at two depths in the wind is illustrated in Figure~\ref{fig_wc4_rad_field}.


\begin{figure}[htbp]
\centering
\includegraphics[width=7.0 cm]{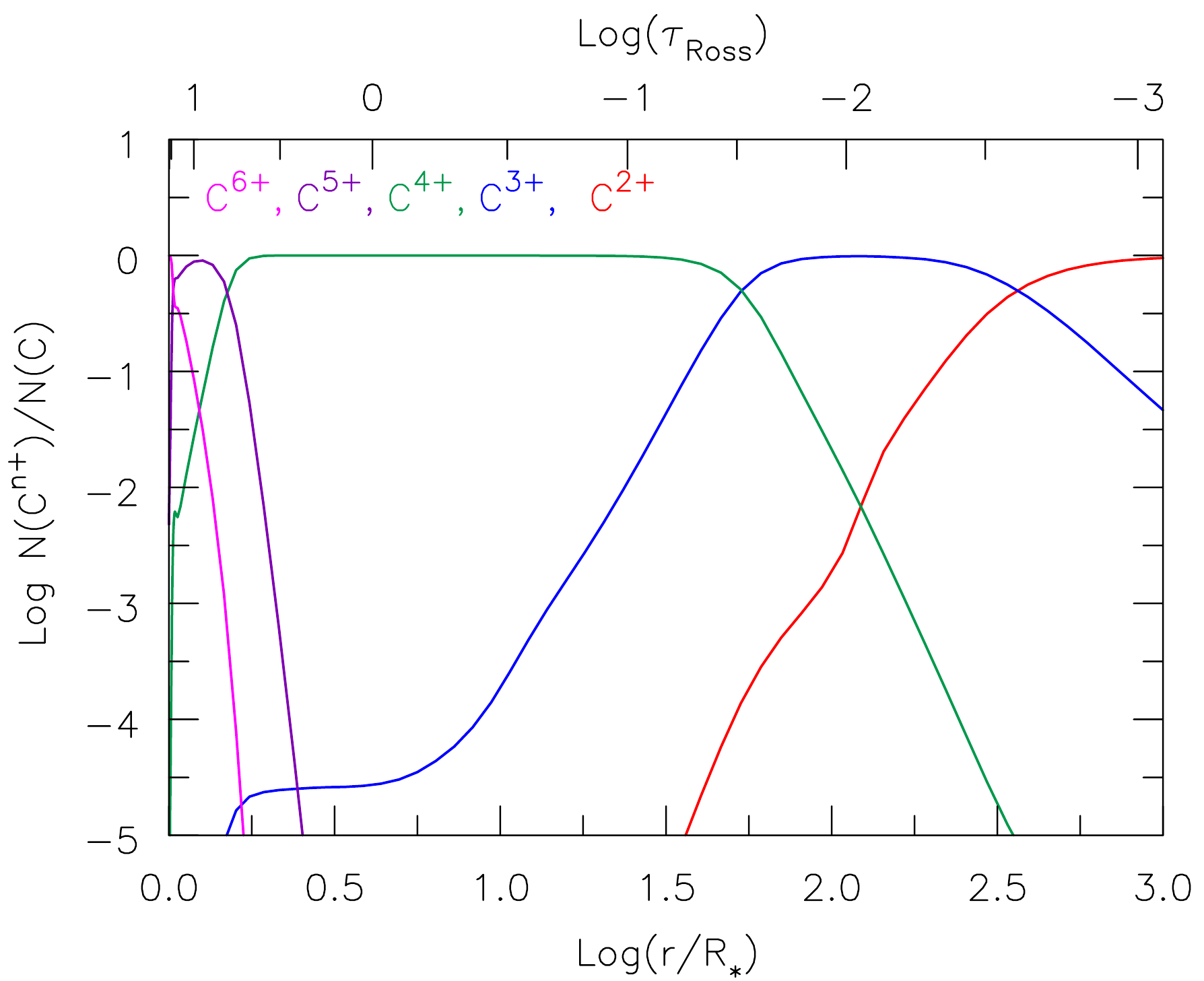}
\includegraphics[width=7.0 cm]{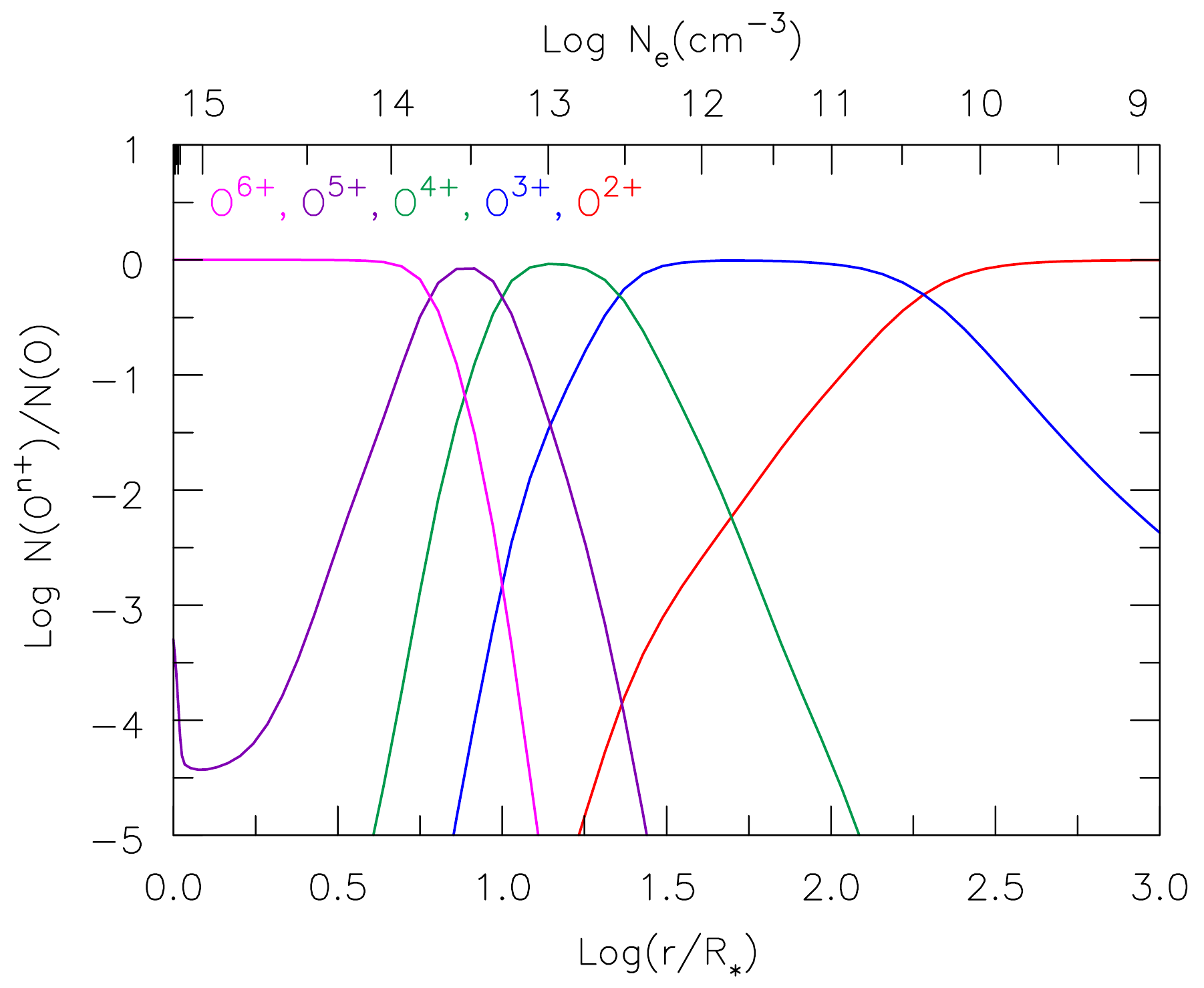}
\caption{Illustration of the stratified ionization structure in the  WC4 star, BAT99-9. On the top
axis we show the Rosseland optical depth (left plot) and the electron density in
units of electrons per cm$^3$ (right plot). The top scales are not linear. As we move out in the wind,
the ionization decreases. This variation is verified by the emission profile line widths -- \ion{O}{6}\
lines are narrower than  \ion{O}{3}\ lines.}
\label{fig_ion_struc}
\end{figure}

A characteristic of WC stars is the stratified ionization structure -- as we move farther out  in the wind
the ionization decreases. The complex ionization structure for C and O is illustrated in 
Figure~\ref{fig_ion_struc}. Because of stratified ionization structure many different species need to
be included to model the spectrum. In BAT99-9 we see emission from four stages of O (\ion{O}{3}
through \ion{O}{6}). To understand driving at the base of the wind additional ionization stages are needed --
in some models we include  \ion{Fe}{4}\ through \ion{Fe}{17}.

The stratified ionization structure is a consequence of several factors. First, the winds of WR stars are not transparent. For example, the \ion{He}{2}\ Lyman continuum (shortward of 228\AA) is optically thick.
Further, the transparency is a strong function of wavelength. Second, as we move out in the wind the radiation field becomes diluted. Third, the intense radiation field in some spectra bands can pump low lying levels. Because of this, and because of the high densities which reduce cascades, ionizations from excited states can play an important role in determining the ionization state of the gas.

\begin{figure}[htbp]
\centering
\includegraphics[width=6.0 cm]{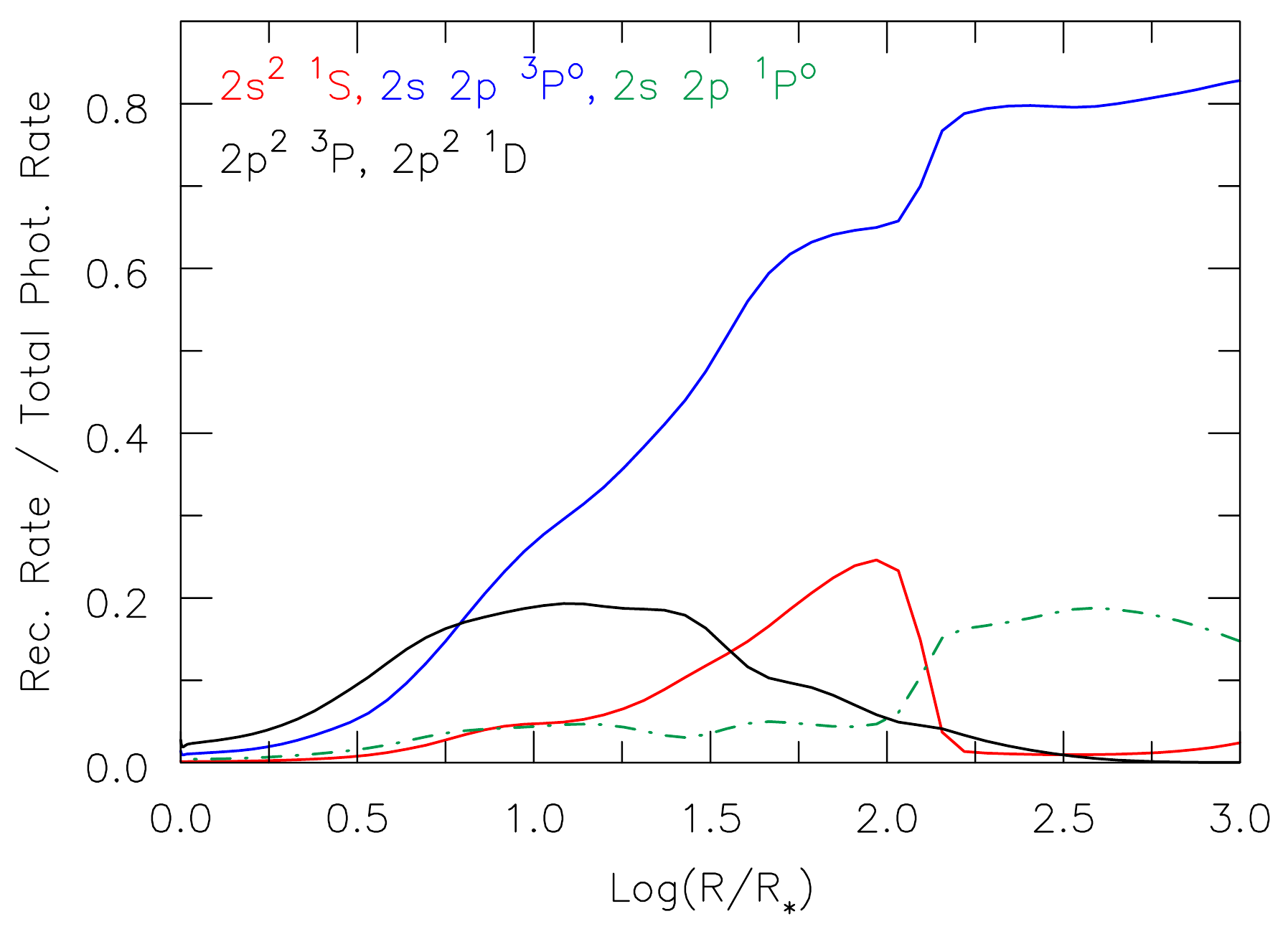}
\includegraphics[width=6.0 cm]{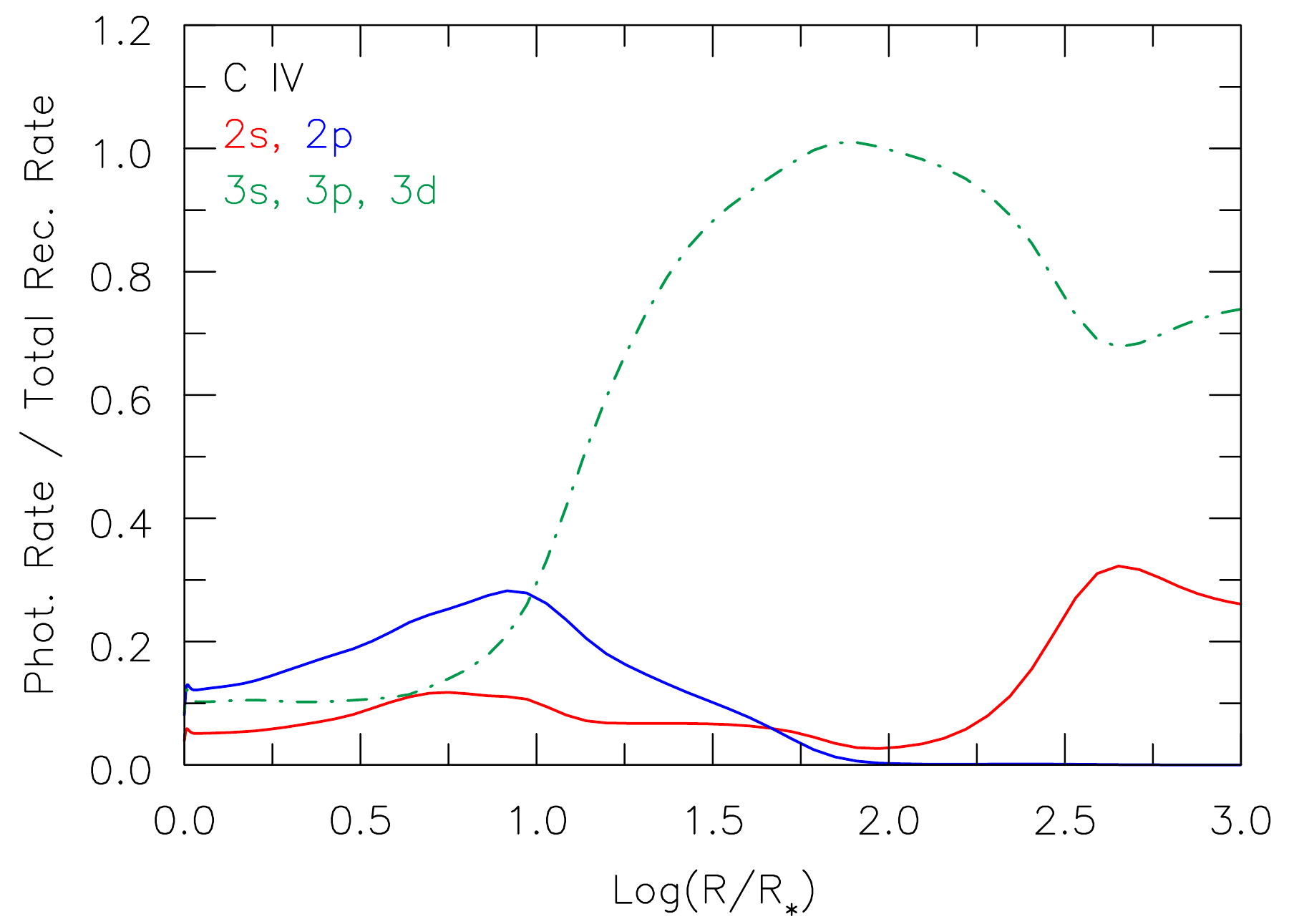}
\caption{Illustration of the model photoionization rates for the  WC4 star, BAT99-9. For illustration
purposes, the rates for the 3{\it l} states have been combined in the right plot.}
\label{fig_phot_rates}
\end{figure}

In  Figure~\ref{fig_phot_rates} we illustrate the photoionization rate, normalized by the total recombination rate to all (included) levels. The normalization was primarily chosen to emphasize the important process in the line formation region. In the inner regions of these dense winds photoionizations and recombinations to each level will be in detailed balance. As we move out in the wind the photoionization from most levels will decrease due to dilution of  the radiation, although for some levels the photoionization and recombination rates may maintain equality if the continua are optically thick.  For \ion{C}{3}\ we see that 3 levels, in order of importance, control the ionization -- 2s\,2p\,\odd{3}{P},   2s$^2$\,\even{1}{S},  2s\,2p\,\odd{1}{P}. On the other hand, for  \ion{C}{4}\ it is the n$=3$ levels (3s, 3p, and 3d) that help to determine the \ion{C}{4}/\ion{C}{5}\ ionization ration. The ionization eventually shifts because the  radiation is becoming diluted (as $1/r^2$) and the populations of the n$=3$ levels are also declining. One reason for the difference in behavior of \ion{C}{3}\ and \ion{C}{4}\  is there is often a rapid decline in the strength of the radiation field shortward of $\sim 228$\AA.

\begin{figure}[htbp]
\centering
\includegraphics[width=6.0 cm]{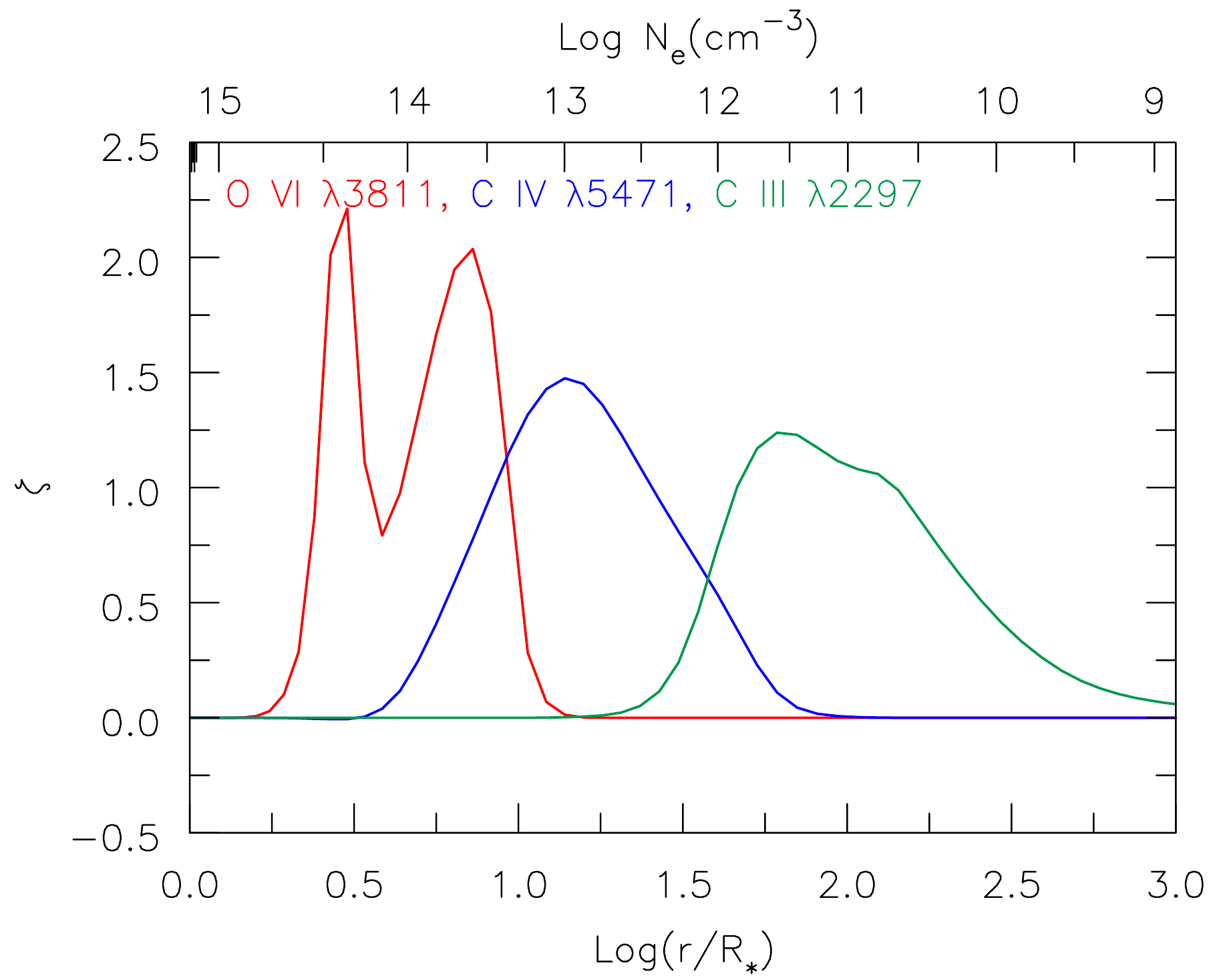}
\caption{Illustration of where  \ion{O}{6}\ $\lambda3811$, \ion{C}{4}\ $\lambda5471$ and
\ion{C}{3} $\lambda 2297$ originate in the wind. $\int \zeta\,d\ln r$ is proportional to
the emission. High excitation lines originate in the inner denser wind, while lower
excitation lines originate in the outer wind.}
\label{fig_orig}
\end{figure}

The presence of multiple ionization stages in the wind results in emission from multiple ionization
stages.  For example, in the case of BAT99-9 we see  emission 
from 2 ionization stages of carbon (\ion{C}{3} \& \ion{C}{4})\footnote{\ion{C}{2} emission
is also predicted but this is masked by blending with other lines.} and 4  ionization stages of oxygen
(\ion{O}{3} through \ion{O}{6}) with the characteristic line width (after allowance for blending
and for the formation mechanism) decreasing as the ionization increases.  The origin of
one O and two C lines is shown in Figure~\ref{fig_orig}  -- it shows that a given emission line
originates over a range of radii, and that lower ionization features form farther out in the wind.

\subsection{\ion{C}{2} in [WC] stars.}
\label{Sect_C2_WC}

The LMC star J060819.93-715737.4 (hereafter J0608) has an exquisite \ion{C}{2} spectrum -- over 150 lines can be identified in the optical \citep{2020ApJ...888...54M, 2021ApJ...906...31W}. It is classified as a [WC11] star with the [] denoting that it is associated with a  low mass
 star ($< \hbox{\rm a few} \Msun$) rather than the product of the evolution of a massive star.\footnote{The 11 appended to WC denotes the ionization class of the star -- in this case, a spectrum dominated by \ion{C}{2}\ with little evidence for \ion{C}{3}.}
The star is probably devoid of H (the observed spectrum exhibits H emission but these probably arise
in circumstellar material and not in the stellar wind) and the C abundance is substantially enhanced (atmospheric mass fractions of He and C are approximately 0.4 and 0.6 respectively). While the rich \ion{C}{2} spectrum is predominately produced by recombination, it cannot be explained by classical optically thin cascades -- optical depth effects play a crucial role in determining the relative \ion{C}{2} line strengths.

The spectrum of J0608 is similar to the  [WC11] star \cpd\  whose spectrum has been extensively discussed and analyzed \citep{1996A&A...312..167L,1997MNRAS.292...86D,1998MNRAS.296..419D,1998MNRAS.297..999D}.
Those studies show the importance of LTDR in producing the spectrum, and identify several optical lines that arise from autoionizing levels.  The spectrum of J0608 has slightly lower ionization  than \cpd, and has a lower terminal wind speed, and as a consequence provides a more ideal object by which to explore the \ion{C}{2}\ spectrum.
 
\begin{figure}[htbp]
\centering
\includegraphics[width=15.0 cm]{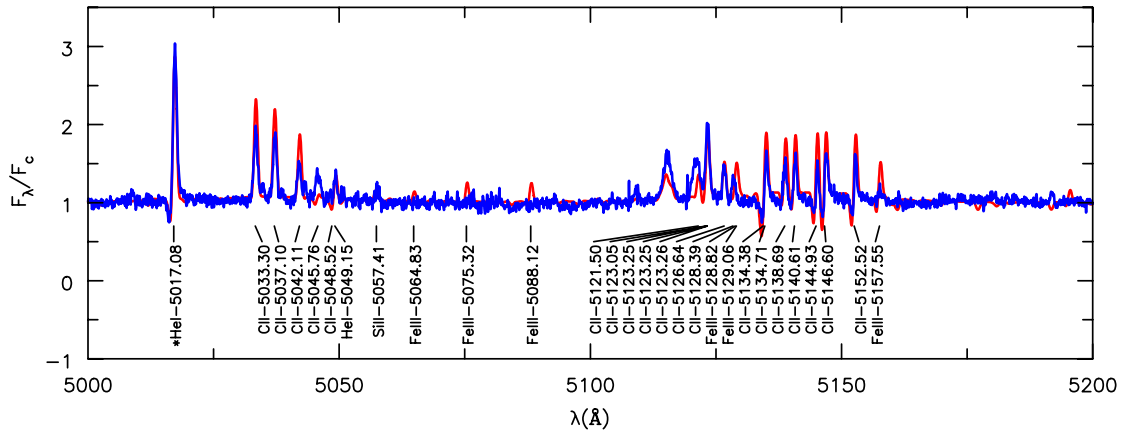}
\caption{A small section of the spectrum of the WC 11 star J0608 (blue) which was obtained by
Nidia Morrell (private communication). A model fit (in red) is shown, and model lines are identified. The broad feature near 5115\,\AA\ (the first unidentified feature) is due to a resonance in the \ion{C}{2} free-free (i.e., Bremsstrahlung) cross-section . It arises from the 2s\,2p(\odd{3}{P})3d\,\odd{2}{P} $-$ 2s\,2p(\odd{3}{P})4f\,\even{2}{D} transition \citep{1993IAUS..155...92B,1998MNRAS.297..999D,2013MNRAS.430..599S}. Both these states are autoionizing.}
\label{fig_cw_wc11}
\end{figure}

 
A small section of the rich \ion{C}{2} spectrum is shown in Figure \ref{fig_cw_wc11}.  \cite{2021ApJ...906...31W} argue that some of the lines are formed via fluorescence processes, but our own modeling suggests that the spectrum can be explained by allowing for the optical depth effects, and by allowing for a transition from ionized to neutral carbon in the outer wind. The latter truncates the
emission of the strongest \ion{C}{3}\ lines.

In Figure \ref{fig_cw_wc11} we also show a free-free resonance ($\lambda \sim$ 5115\,\AA) previously identified in \cpd\ \citep{1998MNRAS.297..999D}. The observations were obtained with a resolution of $\approx\,$7\,\kms, and hence the line is resolved.  In \cmfgen\ the resonance is treated as a free-free resonance since both levels involved in the transition are autoionizing (with $A \sim 10^{11}\,{\rm s}^{-1}$; \cite{1998MNRAS.297..999D}). In the case of this free-free resonance it was trivial to omit it from a ``continuum'' calculation. The latter is needed so we can rectify  the spectrum (i.e., normalize the continuum to unity).
However, this is not the case for bound-free resonances that appear in the photoionization cross-sections. Such resonances can appear in the computed continuum, distorting an otherwise smooth spectrum. These resonances riddle the UV continuum spectrum. However, in practice they are difficult to discern because of the rich forest of bound-bound transitions which mask the continuum spectrum.

\subsection{Supernovae}
\label{Sect_SN}

Supernovae are fascinating objects. They represent the end points of evolution for many stars, and
are an important source of metals (astronomical jargon for all elements more massive than He) in the Universe. Broadly speaking there are two classes of supernovae -- those arising from the core collapse of a massive star \cite[e.g.,][]{2017hsn..book.1053F}, and those arising from the thermonuclear detonation of a white dwarf (WD) star (a compact object of stellar origin with a mass less than $\sim 1.4$\,\Msun\ that is supported by electron-degeneracy pressure). The latter class is designated as a Type Ia SN, and while we know that it involves a WD, we do not know in what type of binary system the explosion occurs.

An extensive discussion of the possible progenitors of Type Ia SNe is given by \cite{2017hsn..book.1151H}. 
Type Ia SN could arise when the WD accretes hydrogen-rich material from a ``normal'' star (e.g. a red supergiant or a main sequence star). As a WD accretes
mass its radius shrinks (and assuming it does not it eject the accreted mass via a surface
explosion in an event called a nova, and which is believed to occur in many systems). As it
approaches the Chandrasekhar mass  of 1.414\,\Msun\ (the upper mass limit or a WD star) it will undergo a thermonuclear explosion. Another possibility is that the WD star accretes He rich material from a WD companion. This material undergoes a surface thermonuclear explosion which triggers an inward propagating shock that triggers the detonation of the accreting WD. A third possibility involves collisions and mergers of two WD stars. In the first scenario the exploding WD has a mass of $~1.4\,\Msun$, which in the other two cases the mass of the exploding WD is (typically) less than $~1.4\,\Msun$. The different scenarios predict different chemical compositions
for the ejecta, and thus determining the chemical composition of the ejecta offers a potential means
of determining the nature of exploding WD.

\begin{figure}[htbp]
\centering
\includegraphics[width=8.0 cm]{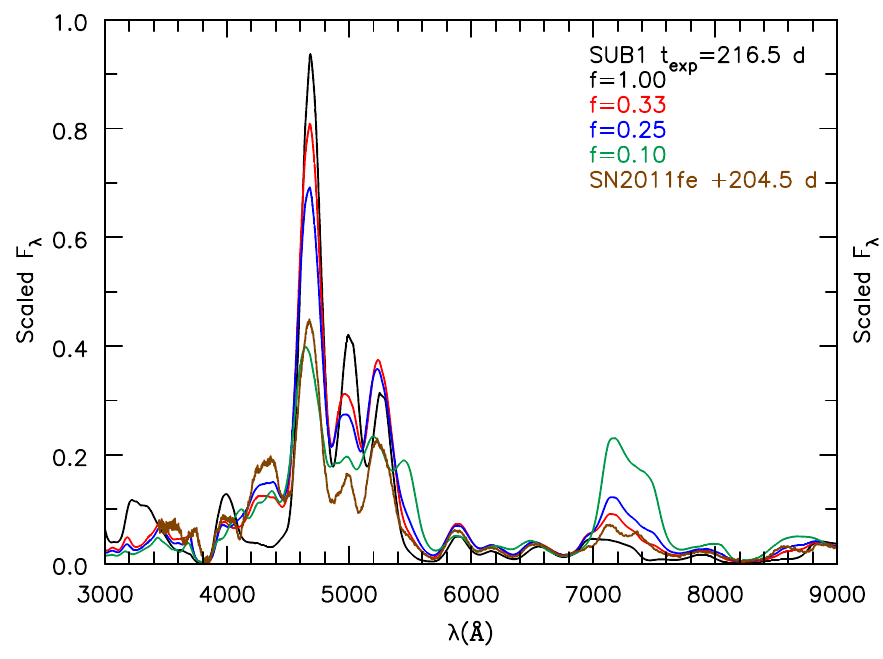}
\caption{Illustration of a small section of the spectrum of a Ia SN showing the influence
of clumping which enhances  recombination, and hence reduces the ionization state of the
gas. In the unclumped model, with $f=1$, \ion{Fe}{3}\ lines near 5000\,\AA\ are too strong.
Increasing the clumping (i.e., decreasing  $f$) lowers the ionization, thus improving the agreement.
However the emission feature near  7400\AA (a blend of \ion{Fe}{2}, \ion{Ni}{2}, and \ion{Ca}{2}  lines)
is worse. The shape of the feature near  5000\,\AA\ is also sensitive to the adopted \ion{Fe}{2}\
photoionization cross-sections, and updated calculations could also improve the fit.}
\label{fig_SNIa}
\end{figure}

At late times (say 200\,days) the spectra of  Type Ia supernovae ejecta are dominated by emission lines of 
Fe, although lines due to Ni, Ca, S, and are also present. One issue with current models of the ejecta is that they fail to yield an iron spectrum in agreement with observation \cite[e.g.,][]{2020MNRAS.494.2221W, 2022MNRAS.512.6150S}.
 Basically, the \ion{Fe}{2}\ lines are too weak relative to \ion{Fe}{3} (Figure \ref{fig_SNIa}), and this limits our ability to interpret ejecta observations. Is the issue related to a problem in the ejecta explosion models, is it due to the ejecta being clumped (which enhances recombination and hence lowers the ionization), is it due to issue with the iron atomic data, is it due to problems treating the
 thermalization of high-energy electrons \citep{2022MNRAS.512.6150S}, or are we missing additional physics? Unfortunately in these systems the iron atomic data is of crucial importance, since  \ion{Fe}{2}/ \ion{Fe}{3}\ is of order unity, and the lines of both species  probably form in the same region. Thus a factor of 2 error in the  \ion{Fe}{2}\ recombination rate will make a factor of 2 error in the \ion{Fe}{2}/ \ion{Fe}{3}\  ionization fraction and will change the relative line strengths (which are produced via collisional excitations) by a factor of 2. Fortunately, for \ion{Fe}{2}\ and  \ion{Fe}{3}, HTDR is unimportant at the relevant temperatures, as can be gleaned from the rates provided by   \cite{1982ApJS...48...95S}.

\section{Conclusion}
\label{Sect_conc}

Accurate photoionization cross-sections are essential for many areas of astrophysics.  The
required quality varies greatly with the application. The biggest needs are at ``intermediate" densities where nLTE is relevant, and where complex density and radiation processes directly affect level populations.  This is the case for many astrophysical phenomena associated with, for example, stellar winds, 
accretion disks, and supernova ejecta.

\funding{Partial support for the work was provided by NASA theory grant 80NSSC20K0524 and
STScI Grant No HST-AR-16131.001-A. STScI is operated by the Association of Universities for Research in Astronomy, Inc., under NASA contract NAS 5-26555.  }

\dataavailability{\cmfgen, and the atomic data used by \cmfgen, are available at  \href{www.pitt.edu/~hillier}{www.pitt.edu/$\sim$hillier}. This site also contains some older O star models.
Data from supernova calculations can be downloaded from 
\href{https://zenodo.org/communities/snrt/?page=1&size=20}{Zenodo} or requested
from the appropriate author. A grid of spectra are available from the \href{http://npollux.lupm.univ-montp2.fr}{Pollux} data base \citep{2010A&A...516A..13P}. A large grid of \cmfgen\ spectra models has been constructed and
is being made available \cite{2021arXiv210714430Z,2020A&A...643A..88Z}. Hillier will also provide \cmfgen\ models upon request.}

\acknowledgments{The author would like to thank P. J. Storey for extensive discussions on atomic data, and for supplying atomic data on \ion{C}{2} and \ion{C}{3} that was directly used in this review. He would also like to thank the numerous workers who have undertaken extensive atomic data calculations and made their work freely available on the internet.  A special thanks to Nidia Morrell who obtained the high resolution spectrum of J0608.  The invaluable comments made by the referees are also greatly appreciated. This paper has made use of NASA's Astrophysics Data System Bibliographic Services.}

\abbreviations{Abbreviations}{
The following abbreviations are used in this manuscript:\\
\\
\begin{tabular}{@{}ll} 
ADS & Astrophysics Data System \\
ESA & European Space Agency \\
HST & Hubble Space Telescope \\
LTE & Local thermodynamic equilibrium \\
nLTE & non local thermodynamic equilibrium \\
LTDR & Low temperature dielectronic recombination \\
HTDR & High temperature dielectronic recombination \\
LBV & Luminous Blue Variable \\
LS coupling & Total orbital angular momentum (L) is coupled with the total spin (S) \\
NASA & National Aeronautics and Space Administration \\
SN & Supernova \\
WD & White Dwarf \\
WR star  & Wolf-Rayet star  \\ 
WN & Wolf-Rayet star belonging to the nitrogen sequence \\
WC & Wolf-Rayet star belonging to the carbon sequence \\
\end{tabular}}

\reftitle{References}

\blankline




\end{document}